\documentclass[preprint]{elsarticle} 

\journal{Journal of Systems and Software}
\usepackage{lmodern}
\usepackage{amssymb,amsmath}
\usepackage{ifxetex,ifluatex}
\ifnum 0\ifxetex 1\fi\ifluatex 1\fi=0 
  \usepackage[T1]{fontenc}
  \usepackage[utf8]{inputenc}
  \usepackage{textcomp} 
\else 
  \usepackage{unicode-math}
  \defaultfontfeatures{Ligatures=TeX,Scale=MatchLowercase}
\fi
\IfFileExists{upquote.sty}{\usepackage{upquote}}{}
\IfFileExists{microtype.sty}{%
\usepackage[]{microtype}
\UseMicrotypeSet[protrusion]{basicmath} 
}{}
\usepackage{hyperref}
\hypersetup{pdfborder={0 0 0}, breaklinks=true}
\usepackage{longtable,booktabs}
\usepackage{rotating}           
\IfFileExists{footnote.sty}{\usepackage{footnote}\makesavenoteenv{longtable}}{}
\usepackage{graphicx,grffile}
\makeatletter
\def\maxwidth{\ifdim\Gin@nat@width>\linewidth\linewidth\else\Gin@nat@width\fi}
\def\maxheight{\ifdim\Gin@nat@height>\textheight\textheight\else\Gin@nat@height\fi}
\makeatother
\setkeys{Gin}{width=\maxwidth,height=\maxheight,keepaspectratio}
\ifx\paragraph\undefined\else
\let\oldparagraph\paragraph
\renewcommand{\paragraph}[1]{\oldparagraph{#1}\mbox{}}
\fi
\ifx\subparagraph\undefined\else
\let\oldsubparagraph\subparagraph
\renewcommand{\subparagraph}[1]{\oldsubparagraph{#1}\mbox{}}
\fi

\usepackage{afterpage}                
\usepackage[tableposition=t]{caption} 

\usepackage{csvsimple}
\usepackage[capitalize]{cleveref}
\usepackage{etoolbox}         
\newtoggle{paper}
\makeatletter
\makeatletter
\@ifclassloaded{elsarticle}{
\toggletrue{paper}
}{
\togglefalse{paper}
}
\makeatother


\usepackage{setspace}           
\usepackage{xspace}             
\usepackage{paralist}           

\newcommand{\practice}[1]{\emph{``#1''}}
\newcommand{\risk}[1]{\emph{#1}}
\newcommand{\quadrant}[1]{\textbf{#1}}

\newcommand{\RQ}{\emph{How does the adoption of scaling agile framework practices address global software development risks?}}
\newcommand{\GSDRiskCatalog}{GSD Risk Catalog\xspace}
\newcommand{\GSDlong}{Global Software Development\xspace}
\newcommand{\GSD}{GSD\xspace}

\newcommand{\pid}[2]{P#1#2\xspace}
\newcommand{\PA}[1]{\pid{A}{#1}}

\newcommand{\PB}[1]{\pid{B}{#1}}

\newcommand{\transcriptOnePMO}{\PB{1}}
\newcommand{\transcriptTwoPMO}{\PB{2}}
\newcommand{\transcriptFourPMO}{\PB{3}}
\newcommand{\transcriptFivePMO}{\PB{4}}

\newcommand{\transcriptTwoPortfolio}{\PB{5}}
\newcommand{\transcriptThreePortfolio}{\PB{6}}
\newcommand{\transcriptSixPortfolio}{\PB{7}}
\newcommand{\transcriptTenPortfolio}{\PB{8}}

\newcommand{\transcriptOneSCAL}{\PB{9}}
\newcommand{\transcriptTwoSCAL}{\PB{10}}
\newcommand{\transcriptThreeSCAL}{\PB{11}}
\newcommand{\transcriptFiveSCAL}{\PB{12}}
\newcommand{\transcriptSixSCAL}{\PB{13}}
\newcommand{\transcriptSevenSCAL}{\PB{14}}
\newcommand{\transcriptEightSCAL}{\PB{15}}
\newcommand{\transcriptNineSCAL}{\PB{16}}

\usepackage{fp}                 

\FPset{\AuthorCount}{0}         
\FPadd{\AuthorCount}{\AuthorCount}{1}
\FPeval{\beechamPos}{clip(\AuthorCount)}
\newcommand{\beechamAuthorPos}{\beechamPos\xspace}
\FPadd{\AuthorCount}{\AuthorCount}{1}
\FPeval{\clearPos}{clip(\AuthorCount)}
\newcommand{\clearAuthorPos}{\clearPos\xspace}
\FPadd{\AuthorCount}{\AuthorCount}{1}
\FPeval{\lalPos}{clip(\AuthorCount)}
\newcommand{\lalAuthorPos}{\lalPos\xspace}
\FPadd{\AuthorCount}{\AuthorCount}{1}
\FPeval{\nollPos}{clip(\AuthorCount)}
\newcommand{\nollAuthorPos}{\nollPos\xspace}

\newcommand{\numOcucoInterviews}{31\xspace}

\newcommand{\numWKRisks}{53\xspace}
\newcommand{\numAllRisks}{63\xspace}
\newcommand{\numNewRisks}{10\xspace}

\newcommand{\numVernerRisks}{85\xspace}
\newcommand{\numVernerWKRisks}{79\xspace}
\newcommand{\numVernerNewRisks}{six\xspace}
\newcommand{\numVernerAlsoNewRisks}{43\xspace}

\newcommand{\numCompoundRisks}{32\xspace}

\FPset{\Aimpl}{58}
\newcommand{\numCaseAimpl}{\Aimpl\xspace}
\FPset{\AnotSeen}{31}
\newcommand{\numCaseAnotSeen}{\AnotSeen\xspace}
\FPeval{\AfracNotSeen}{round((\AnotSeen/\Aimpl)*100:0)}
\newcommand{\fracCaseAnotSeen}{\AfracNotSeen\%\xspace}
\FPset{\Aseen}{27}
\newcommand{\numCaseAseen}{\Aseen\xspace}
\FPeval{\numCaseAremaining}{clip(\Aseen - 1)}

\FPset{\Bimpl}{57}
\newcommand{\numCaseBimpl}{\Bimpl\xspace}
\FPset{\BnotSeen}{23}
\newcommand{\numCaseBnotSeen}{\BnotSeen\xspace}
\FPset{\Bseen}{34}
\newcommand{\numCaseBSeen}{\Bseen\xspace}
\FPeval{\BfracNotSeen}{round((\BnotSeen/\Bimpl)*100:0)}
\newcommand{\fracCaseBnotSeen}{\BfracNotSeen\%\xspace}

\newcommand{\numCaseBseen}{\Bseen\xspace}
\FPset{\BseenPartialImpl}{22} 
\newcommand{\numCaseBSeenPartialImpl}{\BseenPartialImpl\xspace}
\FPset{\BseenPartialAddr}{7}      
\FPeval{\BfracPartialImpl}{round((\BseenPartialImpl/\Bseen)*100:0)}
\newcommand{\fracCaseBpartialImpl}{\BfracPartialImpl\%\xspace}
\FPeval{\Bremaining}{clip(34 - \BseenPartialAddr - \BseenPartialImpl)}

\usepackage{fancyhdr}           
\usepackage{xcolor}
\usepackage[]{draftwatermark}   
\SetWatermarkText{PREPRINT}
\SetWatermarkColor[gray]{0.9}
\SetWatermarkAngle{45}
\SetWatermarkScale{1}

\makeatletter
\@ifclasswith{elsarticle}{preprint}{%
\pagestyle{fancy}
\newcommand{\preprintfoot}[1]{%
\fancyfoot[l]{\vspace{10pt}\footnotesize{\textcolor{gray}{#1}}}
}
\usepackage[final,inline]{showlabels} 
}{%
\newcommand{\preprintfoot}[1]{}    
\usepackage[inline]{showlabels} 
\usepackage{lineno}             
\modulolinenumbers[5]           
\linenumbers
}
\makeatother

\usepackage[obeyDraft]{todonotes} 
\usepackage{xcolor}
\definecolor{pink}{rgb}{1,.4,.4}
\definecolor{red}{rgb}{1, .2, .1}
\newcommand{\notegen}[2]{\todo[inline]{#1: #2}}
\newcommand{\XXXsb}[1]{\notegen{SB}{#1}}
\newcommand{\XXXjn}[1]{\notegen{JN}{#1}}

\PassOptionsToPackage{unicode=true}{hyperref} 
\PassOptionsToPackage{hyphens}{url}

\date{}

\begin{document}
\todo[inline]{Uncomment 'final' documentclass statement.}
\listoftodos

\preprintfoot{This an authors' preprint.  Please cite as: Sarah
  Beecham, Tony Clear, Ramesh Lal, and John Noll (2020) ``Do Scaling
  Agile Frameworks Address Global Software Development Risks?''
  \emph{Journal of Systems and Software, Special Issue on Global
    Software Engineering.}}

\begin{frontmatter}

\title{Do Scaling Agile Frameworks Address Global Software Development
Risks?  An Empirical Study}
\iftoggle{paper}{%

\author{Sarah Beecham\corref{mycorrespondingauthor}}
\cortext[mycorrespondingauthor]{Corresponding author}
\address{Lero - the Science Foundation Ireland Research Centre for Software, \\
University of Limerick, \\
Limerick, Ireland}
\ead{sarah.beecham@lero.ie}
\par
\author{Tony Clear}
\ead{Tony.clear@aut.ac.nz}
\author{Ramesh Lal}
\ead{ramesh.lal@aut.ac.nz}
\address{School of Engineering, Computer and Mathematical Sciences,\\
Auckland University of Technology,\\
Auckland, New Zealand}
\par
\author{John Noll}
\address{School of Physics, Engineering, and Computer Science,\\
University of Hertfordshire\\
Hatfield, Hertfordshire, UK}
\ead{j.noll@herts.ac.uk}
}{%
\author{Sarah Beecham, Tony Clear, Ramesh Lal, John Noll}
\date{}
}
\begin{abstract}
Driven by the need to coordinate activities of multiple agile development teams cooperating to produce a large software
product, software-intensive organizations are turning to scaling agile software development frameworks. Despite the growing adoption of various scaling agile frameworks,  there is little empirical evidence of how effective their practices are in mitigating risk, especially in global software development (GSD), where project failure is a known problem.

In this study, we develop a GSD Risk Catalog of \numAllRisks
risks to assess the
degree to which two scaling agile
frameworks--Disciplined Agile Delivery (DAD) and the Scaled Agile Framework (SAFe)--address software project risks in GSD.  We examined data from two longitudinal case studies implementing each framework to identify the extent to which the framework practices address GSD risks.

Scaling agile frameworks appear to help companies eliminate or mitigate many traditional risks in GSD, especially relating to users and customers. However, several important risks were not eliminated or mitigated. These persistent risks in the main belonged to the Environment quadrant highlighting the inherent risk in developing software across geographic boundaries. Perhaps these frameworks (and arguably any framework), would have difficulty alleviating, issues that appear to be outside the immediate control of the organization.
\end{abstract}

\end{frontmatter}

\section{Introduction }\label{sec:introduction}
With the widespread adoption of agile methods in companies of all
sizes, there is an increasing
need to scale-up agile development beyond a single co-located
team~\cite{Fitzgerald_2017_Continuous}.  This might involve multiple teams collaborating to produce
different subsystems of a single product, multiple teams producing
separate products that are part of a product family, or scaling agile
development upward in the organizational hierarchy, from teams to
program management to the executive suite~\cite{Dingsoyr_2014_Towards}.  Each of these
cases may also require scaling the geographic distance between teams, or
team members on the same team, to accommodate \GSDlong (\GSD).

Companies engage in \GSD for several reasons, including access to new
markets, varied skill sets, and lower labor rates~\cite{Noll_2010_Global, Ebert_2008_Managing,
Nurdiani_2011_Risk}.  \GSD, however, is accompanied by a \emph{risk
tariff}: software development in a co-located context is already a
complex undertaking; globalization increases this complexity,  due to the ``temporal, geographical
and socio-cultural dispersion of stakeholders (managers, developers, clients...)''~\cite{Chadli_2016_Identifying}, which introduce delays, misunderstandings, and mistrust~\cite{Noll_2010_Global}. 
There are additional technical integration issues also to consider~\cite{Cataldo_2011_Factors}, where the architecture needs to consider cross-site coupling and system inter-dependencies~\cite{Sievi-Korte_2019_Challenges}. It would appear therefore that when it comes to GSD, and ``teams operating over the variety of
boundaries, special attention has to be paid to
leveraging the related risks''~\cite{Smite_2007_Project}.

A number of frameworks have been created to guide organizations through
scaling agile development, including Scrum-of-scrums\footnote{https://www.agilealliance.org/glossary/
  scrum-of-scrums}, Large Scale Scrum (LESS)\footnote{https://less.works},
the Spotify Model\footnote{http://blog.crisp.se/2012/11/14/henrikkniberg/scaling-agile-at-spotify},
Nexus\footnote{https://www.scrum.org/resources/nexus-guide}, Scrum at
Scale\footnote{https://www.scruminc.com/scrum-scale-case-modularity},
Disciplined Agile Delivery (DAD)\footnote{http://www.disciplinedagiledelivery.com},
and the Scaled Agile Framework (SAFe\textsuperscript{®})\footnote{http://www.scaledagileframework.com}.
Each of these has its advocates and critics, but to date, there is
scant empirical evidence as to their efficacy in general, and
specifically in managing risk in various settings, such as \GSDlong (\GSD), where teams are distributed around the world.

This study focuses on \GSD, known to have a specific
set of risks to consider, that are additional to those identified by
Oehmen et al~\cite{Oehmen_2014_Analysis} and Wallace and Keil
\cite{Wallace_2004_Software}.  These include task distribution, knowledge
management, geographical distribution, cultural differences, and
communication infrastructure~\cite{Persson_2009_Managing, Verner_2014_Risks}.  As many organizations are now adopting
agile or hybrid development methodologies~\cite{Noll_2019_How, Tell_2019_What} in
globally distributed organizations~\cite{Marinho_2019_Plan} that are
scaling~\cite{Razzak_2018_Scaling}, we ask, 
\begin{quote}
\RQ{}
\end{quote}
To answer this question, we present an industrial multiple case study~\cite{Verner_2009_Guidelines, Bass_2018_Experience} of the
adoption of two scaling agile 
frameworks--Disciplined Agile Delivery
(DAD)~\cite{Ambler_2011_Disciplined}, and the Scaled Agile Framework
(SAFe)~\cite{Leffingwell_2016_Safe}.

In the next section, we provide background on 
project risks in \GSDlong, and give a short overview of our two scaling
agile frameworks, DAD and SAFe.  Following that, in~\cref{sec:method}, we describe our
methodology, and present our results in~\cref{sec:results}. We discuss the implications of the results in~\cref{sec:discussion}, along with the study limitations, and conclude with a summary of
our study and contribution in~\cref{sec:conclusion}.

\section{Background}\label{sec:background}
In this section, we look to the literature to provide an outline and
define types of risk in global software development projects, and then consider
some of the agile frameworks that support large-scale software
development.  We complete this section with our research question
derived from the need to have a better understanding of the impact
these new frameworks might have on project success or failure.

\subsection{Software Project Risk }\label{sec:software-and-risk}

\begin{figure}
\centering
\caption{Wallace and Keil's risk quadrants~\cite{Wallace_2004_Software}}\label{fig:wk-quadrants}
\includegraphics[width=.8\textwidth]{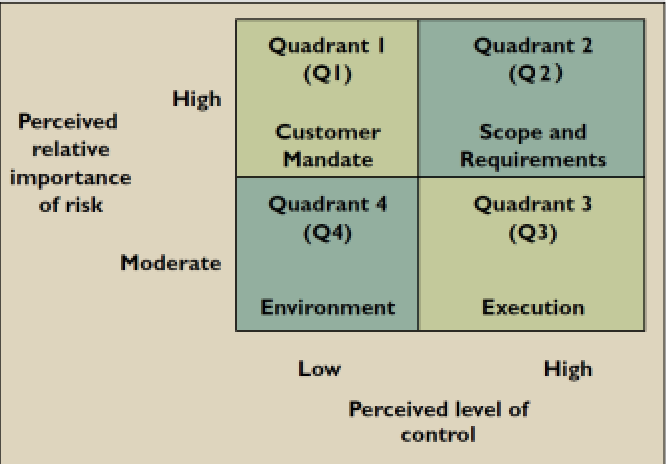}
\end{figure}

The literature on software development risks extends back some years; for instance,
Boehm identified a top 10 list of software project risks in 1991
\cite{Boehm_1991_Software}. In the early 2000s a rigorous set of studies combining 
a Delphi study and survey methods resulted in a comprehensive set of
software project risks~\cite{Wallace_2004_Software, Wallace_2004_Understanding, Schmidt_2001_Identifying}. 
Wallace and Keil~\cite{Wallace_2004_Software}
structured those within a ``risk categorization framework'' which they
derived from the Delphi study and validated through a substantial survey
of over 500 project managers.  That framework comprised \numWKRisks project risks
mapped to the four quadrants shown in \cref{fig:wk-quadrants}.

\begin{figure}
  \centering
  \includegraphics[width=1.1\textwidth]{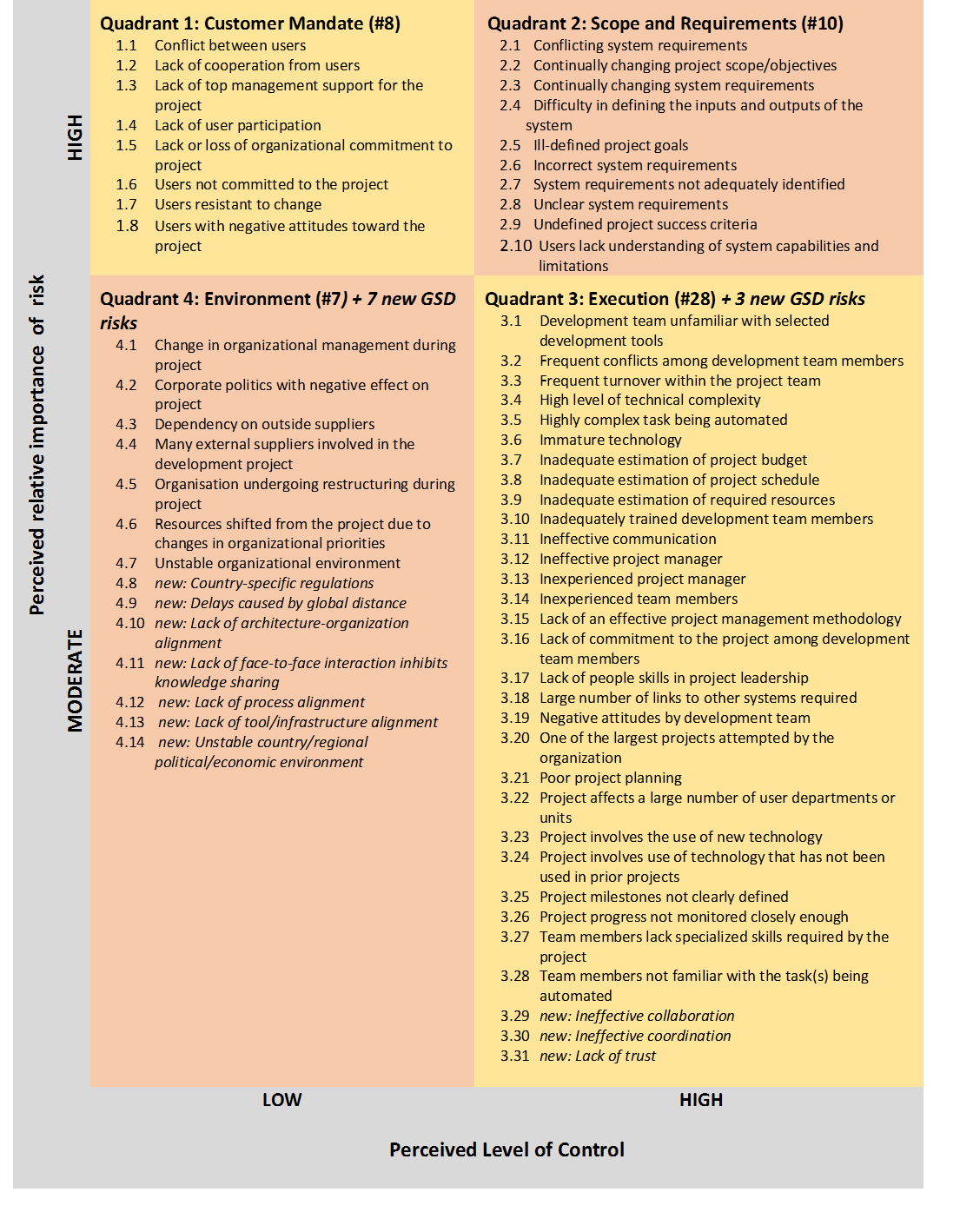}
  \caption{\GSDRiskCatalog derived from Wallace and Keil \cite{Wallace_2004_Software} and Verner et al. \cite{Verner_2014_Risks}}\label{fig:all-risks-quadrant}
\end{figure}

Each of the quadrants in \cref{fig:wk-quadrants} represents a focus on concerns associated with
grouped aspects of risk.  For example, \emph{Customer Mandate} is
perceived as having a high level of importance, and a low level of
perceived control, whereas \emph{Execution} is perceived as being of
moderate importance, with a high level of perceived control. The \quadrant{Customer Mandate} quadrant, for instance, concerns
``risk factors relating to customers and users, including lack of top
management commitment and inadequate user involvement'' \cite{Wallace_2004_Understanding}. The quadrant \quadrant{Scope and
Requirements} involves ``the ambiguity and uncertainties that arise in
establishing the project's scope and requirements''\cite{Wallace_2004_Software, Wallace_2004_Understanding}.
\quadrant{Execution} concerns ``the actual execution of the project\ldots{}
and many of the traditional pitfalls associated with poor project
management''\cite{Wallace_2004_Understanding}.  The \quadrant{Environment} quadrant focuses on
risks associated with the external or internal environment, ``including
changes in organizational management that might affect a
project''\cite{Wallace_2004_Software}. 
The full list of the \numWKRisks risks (plus the \numNewRisks new \GSD risks we add as part of this study) is shown in \cref{fig:all-risks-quadrant}.

Apart from some work by Gotterbarn and Rogerson~\cite{Gotterbarn_2005_Responsible} on ethical risk
analysis, the subsequent focus of attention was on software project
failures, and contributors to software project outcomes~\cite{McLeod_2011_Factors, Lehtinen_2014_Perceived}. As
noted in the framework for understanding influences on software systems
development and deployment project outcomes developed by McLeod and
MacDonell~\cite{McLeod_2011_Factors}, four broad areas of focus were identified:
\emph{institutional context}, \emph{people and action},
\emph{development processes}, and \emph{project content}. Lehtinen et
al.~\cite{Lehtinen_2014_Perceived}, through root cause analysis, followed on the work by
MacLeod and MacDonell to identify the causes and effects of project
failures, adopting a four-element framework of: \emph{people},
\emph{methods}, \emph{task}, and \emph{environment} to guide their
analysis. While not a direct correspondence to Wallace and Keil's four
quadrants
\cite{Wallace_2004_Software}, the coverage and similarities in both of these subsequent
studies are strong, suggesting that the Wallace and Keil framework still
has validity and is a good candidate for evaluating how software risks
are addressed in a scaling agile implementation.  Wallace and Keil's  risk
categorization framework, with its tabulation of risks by quadrant, also
supported detailed and aggregated mapping of risks addressed for each
method (\cref{sec:method}).

Indeed, more recent work in outsourced and global software development
risks~\cite{Abdullah_2009_Outsourced, Verner_2014_Risks}, again drew on the work of Wallace and Keil~\cite{Wallace_2004_Software}.
The outcome was a dual conceptual risk framework for outsourced project
risks comprising the
four elements from Wallace and Keil, and some additional outsourcing
specific risks.\footnote{Wallace et al. (2004) acknowledge that their
  work focuses on in-house software projects, noting that outsourced
  projects may ``exhibit a different risk profile\ldots{}'' and to gauge the risk in these projects may require additional risk
  constructs such as external relationship quality.} The mapping of
global software development risks by Verner and colleagues~\cite{Verner_2014_Risks},
resulted in four risk components from the outsourced risk framework
being identified (\emph{project scope and requirements}, \emph{project
execution}, \emph{project planning and control}, and
\emph{organizational environment}).  A second mapping to the ISO 12207
software lifecycle process, also by Verner and colleagues~\cite{Verner_2014_Risks},
identified \emph{organizational management}, \emph{development process},
\emph{acquisition process}, and \emph{training}--but again was consistent
with aspects of Wallace and Keil~\cite{Wallace_2004_Software}.

In their review Verner and colleagues~\cite{Verner_2014_Risks} identified a broader range
of \GSD-specific risks and mitigations: high level and detailed level \emph{\GSD
vendor selection risks; requirements engineering risks; software
development process risks; architectural design risks; configuration
management risks; culture and social integration risks; training risks;
communication and collaboration risks; planning risks; coordination
risks}, and \emph{control risks}.

Verner et al.'s~\cite{Verner_2014_Risks} set of risks are an
alternative candidate for assessing how well our scaling agile
implementations (in our two case studies) addressed \GSD risks.  Indeed, Verner and
colleagues~\cite{Verner_2014_Risks} suggested that with some
development the Abdullah and Verner outsourcing
framework~\cite{Abdullah_2009_Outsourced}, could be applied as a
``useful framework for \GSD projects.''  Verner et al.'s review
\cite{Verner_2014_Risks} reflects the work of 24 separate studies of
risk in \GSD, including frameworks for distributed software
development threats in Ågerfalk et al.~\cite{Agerfalk_2005_Framework},
and managing risks in distributed software projects in Persson et al
\cite{Persson_2009_Managing}.  Since Verner et al.'s systematic
literature review synthesises the body of work in \GSD risk (resulting
in \numVernerRisks risks), all the risks mentioned are also considered
in our assessment of scaling agile framework's resistance to risk in a
\GSD context (see \cref{sec:method}).  Another candidate set of risks
for GSD was enumerated by Chandli and
colleagues~\cite{Chadli_2016_Identifying}; however, they focus solely
on project management risks, whereas we wanted to take a wider
perspective to include technical risks.  So, since other candidate risk
frameworks were either incomplete or failed to validate their findings
to the same degree as Wallace and Keil, we chose the  Wallace and Keil
framework of four 
quadrants to serve as a frame for incorporating the \GSD related
risks detailed in Verner et al \cite{Verner_2014_Risks}.  The combined view fully described in \cref{sec:results}, and summarised in \cref{fig:all-risks-quadrant}, shows how we augment Wallace and Keil's framework to cater for \GSDlong risks.

Key insights from this combination indicate that the risks identified
in the \quadrant{Customer Mandate} quadrant are not apparent in other
papers on risk we have reviewed here, apart from some coverage in the
``GSD vendor selection''~\cite{Abdullah_2009_Outsourced} grouping, and
the more recent distributed agile development risk observation of
``requirements conflicts amongst multiple product
owners''~\cite{Shrivastava_2017_Risk}.  In keeping with this
observation, Verner et al., in their tertiary review, criticise the
limited focus on developer and vendor perspectives and observe that
``the client is pretty much ignored'' \cite{Verner_2014_Risks} in the
literature.  Perhaps the vendor perspective leads to a proxy customer
situation, which historically has been generally well aligned within a
vendor organisation; however, in a distributed agile setting this can
lead to conflict between product owners~\cite{Razzak_2018_Scaling}.
In our aim to create a comprehensive risk catalog, we found no
additional risks relating to quadrants capturing \quadrant{Customer
  Mandate} and \quadrant{Scope and Requirements}, that are both
thoroughly covered by Wallace and Keil \cite{Wallace_2004_Software}.
Unsurprisingly, the gaps were observed in the \quadrant{Environment},
and \quadrant{Execution} quadrants.  In our \GSD risk mapping
exercise, we were able to categorise all of the risks (or threats)
listed in Verner et al. \cite{Verner_2014_Risks} according to the
Wallace and Keil~\cite{Wallace_2004_Software} four quadrants, despite
many of the \GSD risks appearing at different levels of granularity,
and several presented as compound risks.  The new \GSD risks are
\risk{Delays caused by global distance, Lack of
  architecture-organization alignment, Lack of face-to-face
  interaction inhibits knowledge sharing, Lack of process alignment,
  Lack of tool/infrastructure alignment}, and \risk{Unstable
  country/regional political/economic environment}, all of which fall
under the \quadrant{Environment} quadrant.  We also have three new
risks in the \quadrant{Execution} quadrant, namely \risk{Ineffective
  collaboration, Ineffective coordination} and \risk{Lack of trust.} \cite{Beecham_2020_TR}

In summary, the overall degree of commonality of the other frameworks
with the Wallace and Keil framework~\cite{Wallace_2004_Software} was
sufficient to favor this more established framework for our
comparison augmented with new GSD specific risks from Verner et al \cite{Verner_2014_Risks}. 

\subsection{Scaling Agile Frameworks}\label{sec:agile-scaling-frameworks}

In response to the difficulty of introducing traditional agile methods
(originally designed for small co-located teams) into large-scale
projects and organizations~\cite{Dyba_2008_Empirical}, several scaling agile frameworks
have emerged. These frameworks attempt to scale agile practices for
enterprise-wide agility, to include agile for distributed teams, large
projects and critical systems~\cite{Ebert_2017_Scaling}. Indeed, twenty such frameworks
were identified by Uludağ et al.~\cite{Uludag_2017_Investigating}, of which the Scaled Agile
Framework and DAD frameworks are some of the more popular models
(according to the 12\textsuperscript{th} annual state of agile report
\cite{VersionOne_2018_12th}).

There are debates over the precise definition of ``large scale
agile,'' with Dikert et al.~\cite{Dikert_2016_Challenges} observing
that ``what is seen as large-scale depends very much on the context
and the person defining it.''   Kalenda et al.\cite{Kalenda_2018_Scaling}, distinguish between ``large scale''
and ``very large scale'' and consider a number of further aspects of
scale.  However, in our study, we opt for the
simplicity and comparability of a definition based on number of
people, teams and locations. Therefore, building on the definition of
``large scale'' proposed by Dikert et al. 
\cite{Dikert_2016_Challenges}, we include the additional
stipulation for \emph{large scale agile \GSD} that the definition  explicitly incorporates 
a global focus and thus represents:
\emph{``software development organizations with 50 or more people or at least six
teams''\cite{Dikert_2016_Challenges} and these people or teams must work across sites located in at
  least two different countries}. 
  Development companies, therefore, follow a geographically separated country or company sourcing strategy
according to Vizcaino et al.'s \GSD ontology \cite[p. 74]{Vizcaino_2016_Validated}.

Our research is guided by the observation that ``scaling isn't easy;
large projects often are globally distributed and have many teams that
need to collaborate and coordinate''~\cite{Ebert_2017_Scaling}.
In their column on Scaling Agile~\cite{Ebert_2017_Scaling}, Chris Ebert and Maria Paasivaara
note that there is little support in the empirical literature on large-scale agile practice transformation.

The two scaling agile frameworks in our study both consider risk.
In their introduction to Disciplined Agile Delivery, Ambler and Lines
note: ``The Disciplined Agile Delivery (DAD) process framework
  is a people-first, learning-oriented hybrid agile approach to IT
  solution delivery. It has a \emph{risk-value} life cycle, is
  goal-driven, and is enterprise aware''
~\cite[emphasis added]{Ambler_2011_Disciplined}.
In a similar vein, the Scaling Agile Framework (SAFe), places a great
emphasis on risk mitigation. For example, on the first day of SAFe's Program Increment
(PI) Planning ceremony \footnote{``\emph{Program Increment (PI)}
  \emph{Planning} a cadence-based, face-to-face event that serves as
  the heartbeat of the
  \href{https://www.scaledagileframework.com/agile-release-train/}{{Agile
      Release Train (ART)}}, is a very large meeting in which  all
  teams (remote or collocated) align on the ART to a shared mission
  and \href{https://www.scaledagileframework.com/vision/}{{Vision}}''~\url{https://www.scaledagileframework.com/pi-planning/}}
``teams identify \emph{risks} and dependencies and draft their initial
team PI objectives'' (emphasis added).
SAFe stipulates the importance of PI Planning, by saying, ``if you are
not doing it [PI Planning], you are not doing SAFe.~\footnote{https://www.scaledagileframework.com/pi-planning/}''

\subsubsection{Disciplined Agile Delivery (DAD)}\label{sec:disciplined-agile-delivery}

Disciplined Agile Delivery (DAD) is summarised in the review by Alqudah
\& Razali~\cite{Alqudah_2017_Key, Alqudah_2016_Review} as comprising a set of roles, practices, and
phases. The full details of the method created by Scott Ambler are given
in the book~\cite{Ambler_2012_Disciplined} and the website for the method
\footnote{\url{http://www.disciplinedagiledelivery.com/}}, where the authors
state that \emph{``the Disciplined Agile process decision framework
provides light-weight guidance to help organizations streamline their
processes in a context-sensitive manner, providing a solid foundation
for business agility}.''

We saw our case study site's implementation of DAD unifying the four
levels of the enterprise, encompassing business and software engineering
functions (product management, portfolio management, program management,
and project management). DAD refers to its framework as a toolkit allowing
the creation of an organisation-specific scaling agile approach
regardless of the organisation's size. \emph{``DAD adopts practices and
strategies from existing sources and provides advice for when and how to
apply them together.~ In one sense methods such as Scrum, Extreme
Programming (XP), Kanban, and Agile Modeling (AM) provide the process
bricks and DAD the mortar to fit the bricks together effectively''}
\footnote{\url{http://www.disciplinedagiledelivery.com/}}.

However, central to the DAD framework is portfolio management, which
provides the ability to make the right decisions the first time, removing the
bias in decision making when identifying specific business value ideas
to be pursued either for development as products, features, or for
further experiments, to get feedback and certainty. The governance of
the DAD framework comprises a set of mandated practices spanning the
entire functional setup involved in identifying, developing, and making
software available for use. Hence, portfolio management is driven based
on development and operations intelligence (metrics) including required
guidance and suggestions from the other functional units. The other
major functional process areas contributing to portfolio management are
the product management, and enterprise architecture (including the IT
governance group) within the organisation. Therefore, portfolio
management can make informed decisions in terms of budget and resources,
i.e. their development capacity to be successful in the marketplace.

With the DAD framework, a portfolio can be delivered through various IT
delivery approaches (based on the size of the organisation) such as
program management, agile delivery, continuous delivery, or lean
delivery. There is portfolio management guidance and support for IT
delivery regardless of the delivery means. However, the program
management approach is driven through several projects. While there are
mandated practices spanning the functional setups, portfolio management
emphasizes team autonomy and self-organisation within a cohesive set of 
practices\footnote{\url{https://www.overleaf.com/project/5e7f63a7c581ea00010f6dcf}}.
The DAD framework identifies a set of primary and secondary 
roles and their responsibilities regardless of IT delivery approach,
comprising leadership and more technical roles, and encouraging the
development of ``T-skilled engineers''~\cite{Alqudah_2017_Key}.

DAD allows for formal people management processes
\footnote{\url{http://www.disciplinedagiledelivery.com/\%20agility-at-scale/people-management/}},
   a missing element in many scaling
agile frameworks, although as noted in an earlier study~\cite{Lal_2017_scaling}, we did
not see evidence of those processes in action. While DAD is driven by
agile values and principles, and therefore does not advocate ``big
upfront design,'' in reality the process of managing work item lists from
conception to readiness for development encompasses a large amount of
massaging of functionality and architectural and design work prior to
entering the project level phases. At project level three phases are
incorporated: inception (scoping and sprint planning), construction
(agile testing and coding), and transition (readiness for deployment or
DevOps).

\subsubsection{Scaled Agile Framework (SAFe)}\label{sec:scaled-agile-framework-safe}

SAFe was released in 2011 by Dean Leffingwell and is a living framework
that is continually updated, having (at the time of writing this paper
in 2020) reached version 5.0. According to Leffingwell et al., ``SAFe
applies the power of Agile but leverages the more extensive knowledge
pools of systems thinking and Lean product development''~\cite{Leffingwell_2016_Safe}.
According to case studies featured on the Scaled Agile Framework
website\footnote{www.scaledagileframework.com} (which admittedly may present a one-sided view), SAFe offers many business benefits, including:
\begin{compactitem}
\item 20--50\% increase in productivity,
\item 50\%+ increases in quality,
\item 30--75\% faster time to market,
\item Measurable increases in employee engagement and job satisfaction.
\end{compactitem}

However, these benefits may not be universal, or may come at a cost to work satisfaction amongst the teams~\cite{Paasivaara_2017_Adopting}.
In her study, Paasivaara reports that one of the teams experienced most changes as negative, where 
``teams felt lack of autonomy, as they could no longer decide some things on their own, such as the sprint length. 
With fixed increments they felt moving backward, towards the old waterfall.''
Passivaara concludes that perhaps this negative attitude is due to teams being new to SAFe and they may not have had time to witness the benefits.
This is in contrast to the other case in Paasivaara's study, where all participants described SAFe adoption as ``highly successful.''
The perceived lack of autonomy, and its effects on staff morale, was also found in Noll et al.'s study of SAFe, where there needs to be a balance between autonomy, the feeling of relatedness (support and trust among colleagues), and ability and skills to take on new tasks (following a self-determination theory)~\cite{Noll_2017_Motivation}.  

SAFe offers a soft introduction to the agile world through specifying a
range of structured patterns often needed when organizations transition
from a more traditional environment, particularly in the context of a
large project~\cite{Dyba_2008_Empirical}.
SAFe considers the whole enterprise, and is organized according to
four levels of the organization: Team,
Program, Large Solution\footnote{The ``large solution'' level in SAFe v. 5.0 was termed ``value stream'' in version 4.0; this is the term used in our
SAFe case study}, and Portfolio.
Each level integrates agile and lean practices,
manages its own activities, and aligns with the other levels. Also,
depending on the size of the operation, and the stage the company is at
in terms of scaling, there are varying levels of complexity. For
example, organizations can start with entry-level ``Essential SAFe''
(with just two organizational levels represented--Team and Program).
This can build to ``Large Solution SAFe'' (with an additional level of
the Large Solution), to ``Portfolio SAFe'' (in which the Portfolio level
replaces the Large Solution level). Finally, there is ``Full SAFe'' in
which all four levels of Team, Program, Large Solution, and Portfolio are
represented.

The \emph{Team} level outlines techniques similar to those used in
standard Scrum, with two-week sprint cycles. As in Scrum, teams of
between 5-9 members contain three roles: the Product Owner, Scrum Master,
and team member. Each agile team is responsible for defining, building,
and testing stories from its team backlog in a series of iterations.
Teams have common iteration cadences and synchronization to align their
activities with other teams so that the entire organization is iterating
in unison. Teams use Scrum, XP, or Kanban to deliver prototypes every two
to four weeks~\cite{Leffingwell_2016_Safe}. Important for scaling, all SAFe teams form part
of a team of agile teams--called an Agile Release Train (ART)--that
aims to deliver a continuous flow of incremental releases of value.

At the \emph{Program} level, SAFe extends Scrum using the same ideas but
on a higher level. This level defines the concept of an agile release
train (ART), which is analogous to sprints at the Team level, but
works at a different cadence on a larger timescale.  The  ART is
composed of five sprint cycles.  There is also a sixth ``innovation
planning sprint,'' which allows teams to innovate, inspect, and adapt.
Teams, roles, and activities are organized around the ART~\cite{Leffingwell_2016_Safe}.

Existing roles are stretched and new roles created to cater for the new
responsibilities and practices, where a Product Manager serves as the content authority for the ART, and is
accountable for identifying program backlog priorities. The Product
Manager works with the Product Owners (POs) to optimize feature delivery
and direct the work of POs at the team level. SAFe sees the emergence of
the role of a Release Train Engineer (RTE) ``who facilitates Program-level processes and execution, escalates impediments, manages risk, and
helps to drive continuous improvement''~\cite{Leffingwell_2016_Safe}. Creating this role was
a key success criterion in the case study described by Ebert and
Paasivaara~\cite{Ebert_2017_Scaling}.

The \emph{Large Solution} (previously \emph{Value Stream}) level is
optional, depending on the size of the organization; in larger
organizations, implementation of this level calls for a value stream
engineer (VSE) who plays a similar role to the RTE by facilitating and
guiding the work of all ARTs and suppliers. Leffingwell et al.~\cite{Leffingwell_2016_Safe}
describe these and further roles such as Business Owner, DevOps team
member, Release Manager, and Solution Manager as important.

A highest level of the SAFe hierarchy (made optional in the more recent 4.6 and 5.0 versions) is the \emph{Portfolio level}. This set of executive-level
processes completes the vertical enterprise view, in which senior management make
strategic decisions, deliver value, and prioritize `epics' that are
filtered down to the program level, where they are decomposed into
features, which in turn are fed to the team level in the form of user
stories.

While there is some early evidence in favor of SAFe and its
adoption~\cite{Paasivaara_2017_Adopting, Noll_2016_Global}, perhaps it is too early to judge its true merits or whether
the promised benefits can be universally enjoyed.
From the current literature, it is unclear how well this lean agile enterprise approach mitigates risk in global software development.

\subsection{DAD and SAFe comparison}

In a recent study of scaling agile strengths, weaknesses,
opportunities, and threats (SWOT) in \GSD, Sinha et al. \cite{Sinha_2020_SWOT}
identified six threats, comprising Lack of face to face communication,
Improper task allocation, Cultural differences, Temporal differences,
Linguistic differences, and Lack of agile coaching for scaling. Under
the weaknesses quadrant, they include a Lack of knowledge sharing.
These are recurrent themes in the agile and \GSD risk literature, all of which
we include in our \GSDRiskCatalog.
 
When trying to decide which of the many scaling agile frameworks to adopt, Diebold and colleagues differentiate between a collection of frameworks to include DAD and SAFe \cite{Diebold_2018_Scaling}.  
They find SAFe to have a low level of flexibility (incorporating practices such as Scrum/Kanban/Lean, with specific XP
practices ``mandated''), whereas DAD has a medium level of flexibility (with practices including Scrum/Lean,and a
mixed set of methods) \cite{Diebold_2018_Scaling}. 
They also suggest that further comparative studies are conducted to help with decision making.

Disciplined Agile Delivery (DAD) and Scaled Agile Framework (SAFe) both draw from a variety of agile and lean practices. However, according to Vaidya's observations on three different scaling agile frameworks~\cite{Vaidya_2014_Does}, an organization's context is what matters  most when deciding on which  framework to adopt and which associated practices provide the desired results. We therefore present two case studies in two different contexts, and apply different Scaling Agile Frameworks, to provide some insight into how risk is mitigated. 


Both frameworks (DAD and SAFe) place great emphasis on value and risk, so it seemed appropriate then to evaluate their efficacy at addressing global software development project risks by seeing how well they covered the software project risks identified in our \GSDRiskCatalog.

With this aim in mind, we set out to address the research question, \RQ{}  By 'address' risk, we will specifically look for how the frameworks either mitigate (reduce) or eliminate the \GSD risk.
 In the next section, we explain our approach to answering this question.

\section{Method}\label{sec:method}
\afterpage{
\begin{singlespace}
\begin{tiny}
\begin{longtable}[t]{p{.965\textwidth}}
  \caption{Example of mapping risks from Verner et al.~\cite{Verner_2014_Risks} to \GSDRiskCatalog risks.}\label{tab:verner-wk-mapping-fragment}\tabularnewline
  \toprule
  \emph{\GSDRiskCatalog risk}/Mapped risk from Verner et al.\tabularnewline
  \endfirsthead
  \toprule
  \emph{\GSDRiskCatalog risk}/Mapped risk from Verner et al.\tabularnewline
  \midrule
  \endhead
      \midrule
\multicolumn{1}{c}{\textbf{Quadrant 1: Customer Mandate}} \tabularnewline
\midrule
\emph{Lack of user participation} \tabularnewline
\midrule
National, organizational, and cultural differences of participants can cause problems like rework, loss of data, confusions, etc. \tabularnewline
Lack of collaboration for RE between distributed stakeholders happens due to differences in culture, language distance and processes \tabularnewline
Application of agile practices causes problems in distributed development because of the degree of interaction between stakeholders and number of face-to-face meetings needed \tabularnewline
Requirements information not properly shared with distributed stakeholders affecting their interaction \tabularnewline

\midrule
\multicolumn{1}{c}{\textbf{Quadrant 2: Scope and Requirements}} \tabularnewline
\midrule
\emph{Incorrect system requirements} \tabularnewline
\midrule
Application of agile practices causes problems in distributed development because of the degree of interaction between stakeholders and number of face-to-face meetings needed \tabularnewline
Collaboration difficulties caused by geographic distance in agile development may cause misunderstandings and conflicts \tabularnewline
Problems caused because team members do not share equal knowledge of the domain \tabularnewline
Lack of a common understanding of requirements leads to problems in system functionality \tabularnewline
Lack of collaboration for RE between distributed stakeholders happens due to differences in culture, language distance and processes \tabularnewline
\midrule
\multicolumn{1}{c}{\textbf{Quadrant 3: Execution}} \tabularnewline
\midrule
\emph{new: Lack of trust} \tabularnewline
\midrule
Collaboration difficulties caused by geographic distance in agile development may cause misunderstandings and conflicts \tabularnewline
Mutual trust is important but hard to obtain and lack of trust causes problems. This can be due to lack of face-to-face interaction, cultural differences, and weak social relations \tabularnewline
Fear about the future of jobs and roles, erodes trust \tabularnewline
Limited face-to-face meetings caused by geographic distance impact trust, decision quality, creativity, and general management; knowledge creation is limited within organization. This may lead to problems in creating collaboration know-how and domain knowledge \tabularnewline
Trust among stakeholders necessary to achieve innovation, flexibility, cooperation, and efficiency in distributed environment. Since often a short life span, important to achieve mutual trust rapidly, but if trust is misplaced, entire organization may suffer \tabularnewline
A vendor with poor relationship management can result in problems such as lack of trust \tabularnewline
\midrule
\multicolumn{1}{c}{\textbf{Quadrant 4: Environment}} \tabularnewline
\midrule
\emph{new: Delays caused by global distance} \tabularnewline
\midrule
Application of agile practices causes problems in distributed development because of the degree of interaction between stakeholders and number of face-to-face meetings needed \tabularnewline
Configuration management problems cause dependency, delay and increased time is required to complete maintenance requests \tabularnewline
Temporal and physical distribution increases complexity of planning and coordination activities, makes multisite virtual meetings hard to plan, causes unproductive waits, delays feedback, and complicates simple things \tabularnewline
Inability to communicate in real time) causes collaboration problems; \tabularnewline
Stakeholders located in different time zones can lead to problems in communicating \tabularnewline
Not tailoring organizational structures to reduce delays in problem resolution causes difficulties and can result in site wars and reduce project cohesion \tabularnewline
Choosing a vendor with a lack of control over a project can result in problems such as cost and schedule overruns; and Poor schedule management \tabularnewline

  \bottomrule
\end{longtable}
\end{tiny}
\end{singlespace}
} 

\begin{figure}
  \begin{center}
    \includegraphics[width=\textwidth]{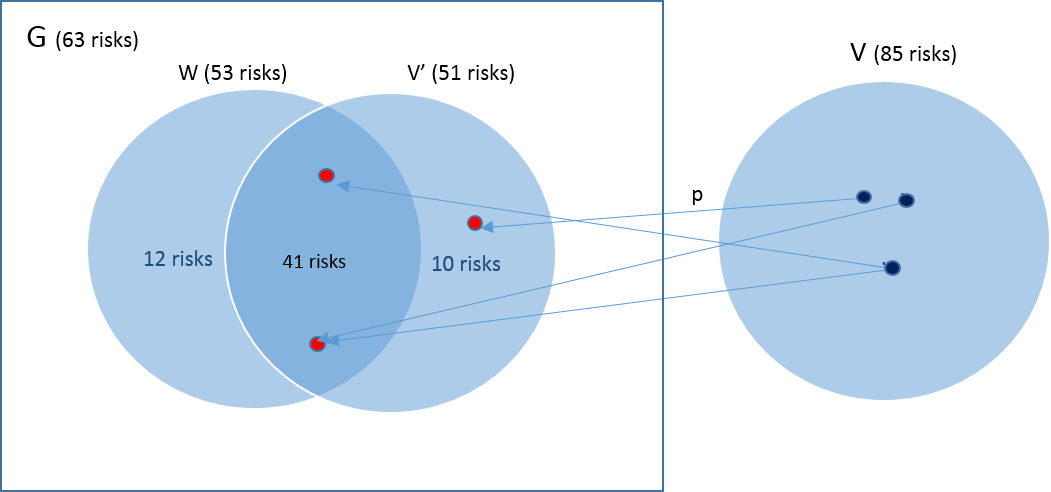}
  \end{center}
  \begin{footnotesize}
    \begin{singlespace}
      In the diagram above, let $G$ be the set of risks in our
      \GSDRiskCatalog,
      $V$ be the set of risks identified by Verner and
      colleagues~\cite{Verner_2014_Risks},
      and 
      $W$ be the set of risks catalogued by Wallace and
      Keil~\cite{Wallace_2004_Software}.  Then,
      \begin{equation*}
        p:{V}\mapsto{W}
      \end{equation*}
      where $p$ is the mapping process we employed to relate risks in Verner
      et al. to risks in Wallace and Keil's catalog, and \emph{$V^\prime$} is the
      subset of risks in the \GSDRiskCatalog to which one or more risks in Verner et al.
      are mapped (i.e. the co-domain of $p$).  The subset of $G$ comprising ten new risks created to
      account for risks from
      Verner et al. that had no, or incomplete, mapping to risks in Wallace
      and Keil's catalog is the relative complement of $W$ with respect to $V^\prime$
      ($V^\prime \setminus W$).
    \end{singlespace}
  \end{footnotesize}
  \caption{\GSDRiskCatalog creation based on Verner et al.~\cite{Verner_2014_Risks} and  Wallace \& Keil~\cite{Wallace_2004_Software}}\label{fig:verner_wk_risk_venn_diagram}
\end{figure}

To answer the research question noted at the end
of the previous section, we take a three-phased approach.  
First we created a catalog of global software development risks we call the \GSDRiskCatalog \cite{Beecham_2020_TR}. 
Second, we created a theoretical mapping of DAD and SAFe practices to the risks in our \GSDRiskCatalog, which we deemed to address each risk; we also rated the degree to which the DAD and SAFe practice(s) eliminated or mitigated the given risk. 
Lastly, we moved from our theoretical model, to a real-world empirical setting in which we examined the extent to which each practice (in DAD and SAFe respectively) was
implemented in a multiple case study comprising two \GSD organisations. This served to illustrate both the extent
to which each scaling agile framework addressed the risks in our catalog,
and how the two case studies implemented the risk-mitigating practices. We examined interview transcripts,
survey results, and observation notes to identify instances where a
risk listed in our 
\GSDRiskCatalog was evident in either of our cases.  From this evidence, we  surmised the extent to which practices in each scaling agile framework addressed \GSD risks.

We describe these three phases in detail in the sequel.


\subsection{Phase 1: GSD Risk Catalog Development}\label{sec:gsd-risk-catalog-method}

Given the context of our study is risk in globally distributed organizations, we first augmented the well-recognised Wallace and Keil \cite{Wallace_2004_Software} set of risks  with additional risks identified in Verner et al.'s~\cite{Verner_2014_Risks} tertiary review of risks in a \GSD context. 

Since there was no one clear, validated set of risks to draw on  specific to \GSD, we took the \numVernerRisks risks detailed in Verner et al.'s review of 24  \GSD risk studies \cite{Verner_2014_Risks} and mapped them onto the established Wallace and Keil risk framework \cite{Wallace_2004_Software}. 
A snippet of this extraction and mapping is shown in \cref{tab:verner-wk-mapping-fragment}.

First, two researchers (author \beechamAuthorPos and author \nollAuthorPos) independently compared \emph{each} risk identified by Verner and colleagues to \emph{each} risk in Wallace and Keil's risk catalog.  If the risk from Wallace and Keil was equivalent to, or would be a consequence of, a risk from Verner et al., we created a correspondence between the two. Mapping the two sets of risks was not straightforward, since some risks in Verner and colleagues' catalog are expressed at a high level, or as a combination of risks; in such cases, the risk from Verner et al was mapped to multiple Wallace and Keil risks.  
Also, we marked those risks in Verner and colleagues' catalog that did not correspond to a Wallace and Keil risk, or was incompletely captured by a set of Wallace and Keil risks, for later consideration.

Second, authors \beechamAuthorPos and \nollAuthorPos reviewed the
independent mapping of the other researcher to establish the level of
agreement; disagreements were discussed between these two authors, and agreement reached.  
Then, author \clearAuthorPos moderated the final result to yield a single unified mapping.
From this mapping, three broad categories emerged: 
\begin{compactenum}
\item a mapped Verner risk, in which a Verner risk was equivalent to one or
more Wallace and Keil risks, 
\item an unmapped Verner risk with no Wallace
and Keil equivalent, and finally 
\item an unmapped Wallace and Keil risk
with no Verner et al equivalent.
\end{compactenum}

In the third step, we coalesced unmapped or partially mapped Verner et
al risks into a small set of new risks.  This was necessary because 
many risks in Verner and colleagues' catalog were at a different level
of abstraction compared to  Wallace and Keil's risks, were
duplicated at the level of abstraction we needed, or appeared to
be more a description of consequences of a risk rather than of the risk itself.

For example, Verner and colleagues identify two risks related to the
rule of law: \risk{Lack of protection for intellectual property rights
  in the vendor country} and \risk{Problems because of differences in
  legal systems such as jurisdiction}.  In the \GSDRiskCatalog, these
rather specific risks coalesced into the more general
\risk{Country-specific regulations} risk.  Similarly, Verner and
colleagues' list \risk{High organizational complexity, scheduling, task
  assignment and cost estimation become more problematic in
  distributed environments as a result of volatile requirements
  diversity and lack of informal communication} and \risk{Lack of a
  common understanding of requirements leads to problems in system
  functionality}; we coalesced these (and ten others) into \risk{Lack
  of face-to-face interaction inhibits knowledge sharing}.

As before, authors \beechamAuthorPos and \nollAuthorPos independently reviewed the candidate \GSDRiskCatalog to double-check that the mappings made sense, and that the new risks were placed in the correct quadrant.  
Again, disagreements were discussed and a resolution agreed between these authors.

Finally, authors \clearAuthorPos and \lalAuthorPos reviewed the
resulting \GSDRiskCatalog to ensure agreement.
\cref{tab:verner-wk-mapping-fragment} shows an extract of the result
(see the companion technical report~\cite{Beecham_2020_TR} for the
full mapping).
The mapping process is depicted in \cref{fig:verner_wk_risk_venn_diagram}.  
As is shown, we added ten risks to Wallace and Keil's inventory,
derived from Verner and colleagues' set of risks that do not map, or only
partly mapped, to the Wallace and Keil risk inventory. 
The remaining 75 risks in Verner et al. corresponded to one or more of
41 of Wallace and Keil's \numWKRisks risks. 
Finally, there were 12 risks from Wallace and Keil's catalog that did not have a
corresponding Verner et al. risk.

In summary, this comparison of risks in Verner et al \cite{Verner_2014_Risks} \GSD 
to the well-established software project risk inventory of Wallace
and Keil~\cite{Wallace_2004_Software} ensured that technical risks as well as project management risks
associated with \GSDlong (\GSD) are considered in our analysis. This comparison resulted in ten new risks
that accommodate risks from Verner et al.'s study that were not
adequately captured by a risk in the Wallace and Keil inventory.
This combined set of \numAllRisks risks, which we label the
``\GSDRiskCatalog,'' is categorised according to Wallace and Keil's
four quadrants: \emph{Customer Mandate}, \emph{Scope and
  Requirements}, \emph{Execution}, and \emph{Environment} (see
\cref{fig:wk-quadrants}), with the majority of new \GSD risks coming
under the \emph{Environment} quadrant.

The full list of \numAllRisks risks in our derived \GSDRiskCatalog is
presented in \cref{fig:all-risks-quadrant}, discussed in
\cref{sec:theoretical-mapping-results}.


\subsection{Phase 2: Theoretical Mapping}\label{sec:theoretical-mapping-method}

Making use of this newly created \GSDRiskCatalog, we then mapped any practice identified in our Scaling Agile Frameworks to the risk, where they appeared to mitigate risk. 
This mapping involved three steps: 

\begin{enumerate}

\item Scaling Agile practice mapped to Risk factors:
  As part of our ongoing, longitudinal case studies, in previous work
  we identified sets of practices from SAFe \cite{Noll_2016_Global},
  and DAD \cite{Lal_2017_scaling, Lal_2018_enhancing}. 
  Four researchers, working in pairs, compared each practice in the scaling agile framework, to each of the \numAllRisks risks
  in the \GSDRiskCatalog (\cref{fig:all-risks-quadrant}).
  Authors \beechamAuthorPos and \nollAuthorPos mapped SAFe practices to the \GSDRiskCatalog,
  and authors \clearAuthorPos and \lalAuthorPos mapped DAD practices to the \GSDRiskCatalog.
  To ensure all researchers worked to the same standard,
  an example of how to `map' a practice to a risk was shared amongst all
  researchers.

\item Strength of mitigation assessment: Once the mapping of practices to risks was completed, each risk was rated according to the degree to which the mapped practices eliminated or mitigated the risk, as to whether the practices ``definitely'' address
  the risk, address the risk ``somewhat'', or do ``not at all''
  address the risk.
  So, if a practice or set of practices unequivocally addressed the risk,
  we coded the practice as ``definitely'';
  if the practice(s) to some extent contributed to elimination or mitigation, 
  we coded the practice ``somewhat''; and when we could not identify a
  practice that would eliminate or mitigate the risk we coded the risk as ``not at all.''
  Again, authors \beechamAuthorPos and \nollAuthorPos assessed SAFe practices, and authors \clearAuthorPos and \lalAuthorPos assessed DAD practices.

\item Inter-rater cross-check (within frameworks): When all
  possibilities were exhausted, each pair of authors reviewed the
  mapping of his or her peer. Any disagreements were discussed within
  each pair until a consensus was reached.

\end{enumerate}

The output from this theoretical mapping was a theory of risk mitigation
according to DAD and SAFe, which is presented in
\cref{tab:dad-mapping} (\cref{sec:dad-risk-mapping}) and
\cref{tab:safe-mapping}  (\cref{sec:safe-risk-mapping}).


\subsection{Phase 3: Empirical Evidence}\label{sec:method-empirical-evidence}

\XXXsb{DONE? we need to explain more about the aim of the empirical studies - which is longitudinal in both cases, and has a broader  aim, - it is a threat that we did not set out to look specifically at risk. We need to give the setting here.  E.g. number interviewed, dates, etc.  We can abstract and give ids etc in results}

In the last stage, we examined data collected in a longitudinal
multiple case study of two companies engaged in scaling agile
development adoption according to the DAD (Case A) and SAFe (Case B)
frameworks. A two case multiple case study has advantages over a
single case, as, according to Yin, ``\ldots{} analytic benefits from
having two (or more) cases may be substantial\ldots{}'' since,
``analytic conclusions independently arising from two cases \ldots{}
will be more powerful than those coming from a single-case \ldots{}
alone''~\cite{Yin_2018_Case}. Furthermore, Yin states that when asking
``How'' and ``Why'' types of questions, case study research is
particularly relevant~\cite{Yin_2018_Case}. Our case boundaries
include time (in years), geographic locations, domain, and practice
adoption. We apply the multiple case study design to address our
``How'' research question, and test our theoretical mapping of scaling
agile framework practices (described in Phase 2 of our method) in
which we hypothesise, that a given set of scaling agile practices can
mitigate \GSD risk. The business model for both cases is to develop,
maintain and sell software to clients throughout the globe. The
original aim of both of our case studies was to gain broad insight
into issues and benefits of scaling agile
framework adoption; as such, while not focused specifically on risk,
these studies yielded a rich source of data from which we were able to
identify many issues related to risk, test our theory, and compare and contrast across cases.

\subsubsection{DAD evaluation--Case A} 
\begin{table}[t]
\centering
\caption{List of participants interviewed in Case A.}
\label{tab:list-of-participants-A}
\begin{tiny}
\begin{tabular}{lp{.31\textwidth}cccp{0.18\textwidth}}
\toprule
\textbf{Part. ID} & \textbf{Role} & {\textbf{Team}} & {\textbf{Program}} & {\textbf{Portfolio}} & \textbf{Country} \\
  \midrule
\PA{1}         & Product Owner/ Product Management                           & 1 & 1 & 1 & Australia           \\
\PA{2}         & Principal Engineer/ Enterprise Architect/ Program Architect & 1 & 1 & 1 & Australia/USA/India \\
\PA{3}         & Software Engineer                                           & 1 &   &   & Australia           \\
\PA{4}         & Quality Assurance Manager                                   & 1 & 1 & 1 & Australia/USA/India \\
\PA{5}         & Senior Software Engineer                                    & 1 & 1 &   & Australia           \\ 
\PA{6}         & Team Lead                                                   & 1 & 1 &   & Australia           \\
\PA{7}         & Engineering Manager                                         & 1 & 1 &   & Australia           \\
\PA{8}         & Director of Software Engineering                            & 1 & 1 & 1 & Australia           \\
  \midrule
\textbf{Total} &                                                             & 8 & 7 & 4                       \\
\bottomrule
\end{tabular}
\end{tiny}
\end{table}

For the empirical investigation of the DAD method, interview
transcripts from Case A (described in \cref{sec:results}) were
examined by authors \clearAuthorPos and
\lalAuthorPos.
In Case A, author \lalAuthorPos 
conducted interviews of participants in March of 2017~\cite{Lal_2017_scaling,
Lal_2018_enhancing}; see \cref{sec:case-study-a} for a description of
the company (``Company A'') involved in Case A, and details regarding
the interview participants (\cref{tab:list-of-participants-A}).
The interviews were conducted at Company A's 
software engineering lab at Box Hill, Melbourne, Australia. A total of
eight interviews (each an hour long) were conducted one week in March 2017. Participants were identified based on the various
software engineering roles within this software vendor organisation.
These roles were: Principal Software Engineer, Senior Software Engineer,
Software Engineer, Team Leader, Engineering Manager, Product Manager or
Product Owner, Quality Assurance Manager, and the Director of Software
Engineering. All the participants for this investigation were part of
their DAD transformation. The interview instruments were based on the
reasons and approach for switching to the DAD method. An agreement was
made to record the interviews that were later transcribed.
Although all
interviews were conducted out of the Melbourne site only, the projects
covered development across Australia, the USA, and India. 

\subsubsection{SAFe Evaluation--Case B} 
\begin{table}[t]
\centering
\caption{List of interviewees from Case B.  Locations and some roles are general, to
  preserve anonymity; \emph{``Sr Management''} represents
  \emph{Architect, Product Manager, Regional Manager}, and {Development Manager} roles} 
\label{tab:list-of-interviewees-B}
{\tiny
\begin{tabular}{lp{.29\textwidth}cccl}
\toprule
\textbf{Part. ID}           & \textbf{Role}                 & {\textbf{Team}} & {\textbf{Program}} & {\textbf{Portfolio}} & \textbf{Location} \tabularnewline
\midrule                                                    
\transcriptOnePMO{}         & Scrum Master, Project Manager & 1               & 1                  &                      & North America \tabularnewline
\transcriptTwoPMO{}         & Scrum Master, Project Manager & 1               & 1                  &                      & UK/Ireland \tabularnewline
\transcriptFourPMO{}        & Sr Management                 &                 &                    & 1                    & Continental Europe \tabularnewline
\transcriptFivePMO{}        & Scrum Master, Project Manager & 1               & 1                  &                      & Continental Europe\tabularnewline
\transcriptTwoPortfolio{}   & Sr Management                 &                 &                    & 1                    & Continental Europe \tabularnewline
\transcriptThreePortfolio{} & Sr Management                 &                 &                    & 1                    & UK/Ireland \tabularnewline
\transcriptSixPortfolio{}   & Sr Management                 &                 &                    & 1                    & UK/Ireland \tabularnewline
\transcriptTenPortfolio{}   & Sr Management                 &                 &                    & 1                    & North America \tabularnewline
\transcriptOneSCAL{}        & Developer                     & 1               &                    &                      & North America \tabularnewline
\transcriptTwoSCAL{}        & Product Owner                 & 1               &                    &                      & North America \tabularnewline
\transcriptThreeSCAL{}      & Scrum Master, Project Manager & 1               & 1                  &                      & North America \tabularnewline
\transcriptFiveSCAL{}       & Sr Developer, Tech Lead       & 1               & 1                  &                      & North America \tabularnewline
\transcriptSixSCAL{}        & Developer                     & 1               &                    &                      & UK/Ireland \tabularnewline
\transcriptSevenSCAL{}      & Developer                     & 1               &                    &                      & UK/Ireland \tabularnewline
\transcriptEightSCAL{}      & Sr Developer                  & 1               & 1                  &                      & UK/Ireland \tabularnewline
\transcriptNineSCAL{}       & Product Owner                 & 1               &                    &                      & UK/Ireland \tabularnewline

  \midrule
\textbf{Total} &  & 11 & 6 & 5 \\
\bottomrule
\end{tabular}
}
\end{table}

\begin{table}[t]
\centering
\caption{List of participants in Case B ``self-assessment'' survey (responses aggregated to preserve anonymity).}
\label{tab:list-of-participants-B}
\begin{tiny}
\begin{tabular}{p{.25\textwidth}cccp{0.33\textwidth}}
\toprule
\textbf{Role}               & {\textbf{Team}} & {\textbf{Program}} & {\textbf{Portfolio}} & \textbf{Country}                            \\
  \midrule
  Project Mgr./Scrum Master & 9               & 6                  & --                   & Canada, France, Ireland, Norway, Poland, UK \\
  Developer                 & 20              & --                 &                      & Canada, Ireland, Denmark                    \\
  Quality Assurance         & 8               & --                 & --                   & France, Ireland, USA                        \\
  Product Owner             & 4               & 5                  & --                   & Canada, Ireland, USA                        \\
  Director of Eng.          & 1               & --                 & --                   & Ireland                                     \\
  Development Manager       & 1               & 1                  & 1                    & Ireland                                     \\
  Product Manager           & 2               & 1                  & 1                    & Spain                                       \\
  Quality Assurance Lead    & --              & 1                  & --                   & Ireland                                     \\
  Database Administrators   & --              & 1                  & --                   & Ireland                                     \\
  Technical Support         & --              & 1                  & --                   & Ireland                                     \\
  Chief Executive Officer   & --              & --                 & 1                    & Ireland                                     \\
  Chief Technology Officer  & --              & --                 & 1                    & Ireland                                     \\
  Regional Management/Sales & --              & --                 & 2                    & France, Italy                               \\
  \midrule
  \textbf{Total}            & 45              & 16                 & 6                    &                                             \\
\bottomrule
\end{tabular}
\end{tiny}
\end{table}

Data relating to SAFe were obtained from results of an ongoing
longitudinal participant-observer study (called ``Case B''), with
moderate researcher involvement, that began at the end of 2015 and
continued through to the autumn of 2019.  Similar to Case A, the main
purpose of our collaboration was to observe how the company (``Company
B'') transitioned
from a plan-driven development process to a scaling agile development
process based on SAFe practices.  Company  B is specifically
interested in how to adopt the new agile, lean, and Kanban practices in
their highly distributed setting: where teams \emph{and} individual team \emph{members} are globally distributed (see \cref{fig:case-map}.

During our four year collaboration, authors \beechamAuthorPos and \nollAuthorPos, along with
their colleagues, conducted \numOcucoInterviews
interviews of team members and managers in a variety of roles, at all
levels of the company, chosen to be representative of all levels of
the development organization\footnote{For the questions asked in our
  semi-structured interviews please see our companion Technical Report
  \cite{Beecham_2017_Lean}}.
Interviews were conducted  on-site in the company's Dublin
headquarters, and via video conference, over two years, starting November,
2015,
Participants of sixteen of these interviews are directly quoted in our study, as listed in \cref{tab:list-of-interviewees-B}.

We also observed distributed development teams conducting Scrum
`ceremonies,' such as daily standups, sprint planning, and
retrospective meetings; and, we observed weekly program-level ``scrum
of scrums'' style meetings.  These observations began in November 2015
and continued, focusing on different teams, until the end of 2017.
Observations helped to place the interviews in context, but did not directly
provide any data for this study.

Finally, a series of three SAFe ``self-assessment'' surveys were
administered to various teams, and program and portfolio level
participants, in February 2017, July 2017, and March 2018
~\cite{Noll_2016_Global, Razzak_2017_Transition, Razzak_2018_Scaling}; see \cref{tab:list-of-participants-B} for
details regarding the participants in the surveys.  The self-assessment surveys identified the level to which the participants perceived they implemented various SAFe practices and ceremonies.

\subsubsection{Within and between multiple case study evaluation} 

As a first step, we examined these data for evidence that the
companies had experienced problems (or not) related to the risks in
our \GSDRiskCatalog

Then, in the second step, we assessed the frequency at which each
company performed the respective scaling agile framework practices
mapped to risks in the \GSDRiskCatalog.  In Case A, this frequency was
assessed to be ``always'' as the company had completed its agile
adoption, at the time the interviews took place.  In Case B, data from
the self-assessment surveys were examined to determine the company's
self-assessed frequency of practice performance.

Finally, working in pairs (authors \clearAuthorPos and \lalAuthorPos for case A, and authors \beechamAuthorPos and \nollAuthorPos for Case B), we connected the output of the previous two
steps, to understand whether the scaling agile practices eliminated or
mitigated the corresponding risks: 
\begin{inparaenum}
\item if the practices were implemented in the company,
and no evidence of the risk was seen, the practices could have been
material in \emph{eliminating} the risk;
\item if the practices were
implemented in the company, but there was evidence that the risk was a
problem for the case, the practices still might have been effective at
\emph{mitigating} or reducing the risk; or, 
\item this might indicate that the
theoretical mapping is not effective in practice.
\end{inparaenum}

To determine which of these alternatives was the case, we considered three additional elements:
\begin{compactenum}
\item \emph{Strength of theoretical mapping:} the degree to which the practices address the risk.  Risks that are only ``somewhat'' addressed (by the practice), are perhaps more likely to be seen as problems.
\item \emph{Strength of practice implementation in cases:} the frequency at which the associated practices were performed.  If this was less than ``always,'' it's possible the practices were not effective because they were not thoroughly implemented.
\item \emph{Level of control:} whether the risk can be \emph{eliminated}, or only \emph{mitigated}.  Certain risks, such as \risk{Unstable country/regional political/economic environment},  are part of the environment; they cannot be eliminated, but their impact can be reduced.
\end{compactenum}

\noindent We present the results of applying this method in the sequel.

\section{Results}\label{sec:results}

In this section we first present a new catalog of \GSD risks created by
comparing and merging risks identified by Verner and
colleagues~\cite{Verner_2014_Risks} to Wallace and Keil's list of
risks~\cite{Wallace_2004_Software}.  Then, we show the theoretical mapping
of SAFe and DAD practices to these \GSD risks.  Finally, we present
empirical evidence of the effectiveness of our theoretical mapping, that underpins the extent two which the scaling agile frameworks eliminate or mitigate risks in \GSD.


\subsection{\GSD Risk Catalog development}\label{sec:gsd-risk-catalog}

To create a comprehensive catalog of risks faced by global software
development projects, 
we compared \numVernerRisks{} risks identified by Verner and
colleagues~\cite{Verner_2014_Risks} in their tertiary study of risks
in global software development, to the \numWKRisks{} in Wallace
and Keil's~\cite{Wallace_2004_Software} risk framework.  We found many risks identified in Wallace and Keil related to GSD risks, and that
many of the risks listed by Verner et al. identify more than
one risk.  For example, a risk listed under ``Requirements
engineering risks and mitigation advice'' states, ``A lack of suitable
tools or methodologies available for requirements elicitation may lead
to problems in obtaining the real requirements.~\cite[Table 9,
p. 64]{Verner_2014_Risks}'' This statement articulates two
risks: a lack of suitable \emph{tools} for requirements elicitation,
and a lack of suitable \emph{methodologies} for requirements
elicitation.  Over a third (\numCompoundRisks) of the risks in Verner
and colleagues' list could be classified as ``compound'' risks of this
nature.

Other risks identified by Verner et al. are general, high-level risks
that could have multiple consequences for a software development
project.  For example, the category ``Software development process risks and
mitigation advice'' includes this risk:
``Application of agile practices causes problems in distributed
development because of the degree of interaction between stakeholders
and number of face-to-face meetings needed.~\cite[Table 10, p. 64]{Verner_2014_Risks}'' 
This high-level risk leads to several risks identified by Wallace and
Keil, including:
Lack of user participation (quadrant 1),
Incorrect system requirements (quadrant 2), and
Inadequate estimation of project budget (quadrant 3).

This high-level risk also leads to several new, \GSD-specific risks not
found in Wallace and Keil's catalog, including:
Delays caused by global distance,
Ineffective collaboration,
Ineffective coordination, and
Lack of face-to-face interaction inhibits knowledge sharing.
Of the \numVernerRisks risks identified by Verner and colleagues,
\numVernerWKRisks correspond to, imply, or result in at least one risk in Wallace and
Keil's catalog.  

We also found \numVernerNewRisks that had no
correspondence to any of Wallace and Keil's risks:
\begin{compactenum}
\item ``Lack of well-defined modules causes problems with progressive integration~\cite[Table 11, p. 65]{Verner_2014_Risks},''
\item ``Project participants have limited understanding of other project participants' competencies~\cite[Table 15, p. 67]{Verner_2014_Risks},''
\item ``Coordination in multi-site development becomes more difficult in terms of articulation of work as problems from communication lack of group awareness and complexity of the organization appear and influence the way the work must be structured~\cite[Table 17, p. 68]{Verner_2014_Risks},''
\item ``Tool compatibility may prove a problem; sites are likely to prefer different programming languages, support tools, operating systems, and development tools~\cite[Table 18, p. 69]{Verner_2014_Risks},''
\item ``No process alignment, in terms of traditions, development
  methods, and emphasis on user involvement, will often differentiate
  between sites, possibly resulting in incompatibility and
  conflicts~\cite[Table 18, p. 69]{Verner_2014_Risks},'' and
\item ``A vendor with poor relationship management can result in problems such as lack of trust~\cite[Table 8, p. 63]{Verner_2014_Risks}.''
\end{compactenum}

And, we found \numVernerAlsoNewRisks risks that, due to being high-level or
compound in nature, not only corresponded to one or more risks in
Wallace and Keil's catalog, but also suggested a new risk, not in
Wallace and Keil's catalog.  As a consequence, we formulated
\numNewRisks additional risks; these are listed in \cref{tab:new-risks}.
\begin{small}
\begin{singlespace}
\begin{table}
\centering
\caption{New risks created to augment Wallace and Keil's
  catalog, and corresponding quadrant in Wallace and Keil's framework
  (quadrant 3 is ``Execution,'' quadrant 4 is ``Environment'').}\label{tab:new-risks}
\begin{tabular}{lc}
\toprule
New risk                                                    & Quadrant \tabularnewline
\midrule
Country-specific regulations                                & 4 \tabularnewline
Delays caused by global distance                            & 4 \tabularnewline
Ineffective collaboration                                   & 3 \tabularnewline
Ineffective coordination                                    & 3 \tabularnewline
Lack of architecture-organization alignment                 & 4 \tabularnewline
Lack of face-to-face interaction inhibits knowledge sharing & 4 \tabularnewline
Lack of process alignment                                   & 4 \tabularnewline
Lack of tool/infrastructure alignment                       & 4 \tabularnewline
Lack of trust                                               & 3 \tabularnewline
Unstable country/regional political/economic environment    & 4 \tabularnewline
\bottomrule
\end{tabular}
\end{table}
\end{singlespace}
\end{small}

The result of combining the \numNewRisks new risks with the
\numWKRisks{} in Wallace and Keil's catalog,  yielded a combined \GSDRiskCatalog
of \numAllRisks{} risks, illustrated in \cref{fig:all-risks-quadrant}.
The full correspondence
of risks identified by Verner and colleagues, to risks in the
\GSDRiskCatalog, is available in a companion technical report~\cite{Beecham_2020_TR}.


\subsection{Theoretical Mapping of Scaling Agile Practices to \GSD risks}\label{sec:theoretical-mapping-results}

To understand the extent to which scaling agile frameworks address \GSD risks, we
assessed how well practices in DAD and SAFe address the risks in our
\GSDRiskCatalog.  The result is a \emph{theoretical} mapping of scaling
agile practices to \GSD risks.

\cref{tab:dad-mapping}  (\cref{sec:dad-risk-mapping}) and \cref{tab:safe-mapping}  (\cref{sec:safe-risk-mapping}) show our assessment
of the degree to which DAD and SAFe practices (respectively) address
the risks in our \GSDRiskCatalog, 
along with examples of 
DAD and SAFe practices that address the given risk. Space does not
allow all the associated DAD and SAFe practices to be included, so the tables
present selected examples; the complete mapping is available as part of a technical report~\cite{Beecham_2020_TR}.

\begin{figure}
  \centering
  \includegraphics[width=\textwidth]{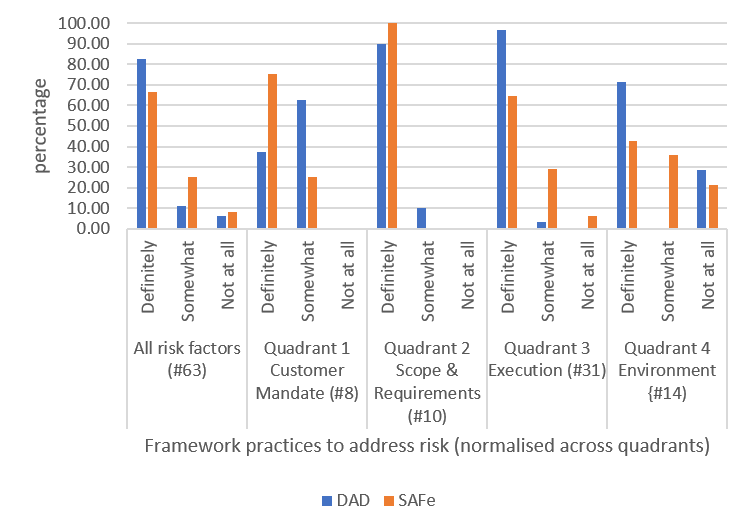}
  \caption{Comparative strength of risk mitigation across all \numAllRisks{} factors}\label{fig:mitigation-all}
\end{figure}

\cref{fig:mitigation-all} summarizes the extent to which each
scaling agile
framework theoretically addresses the \GSDRiskCatalog risks. Looking at the total,
\cref{fig:mitigation-all}  shows that both frameworks address most of
the risks.  The raw figures have been normalized across quadrants to allow a comparison of how each framework addresses 
risks in the \GSDRiskCatalog quadrants outlined in \cref{fig:all-risks-quadrant}.  Despite the frameworks addressing a similar number of risks, we see some differences when we compare risk mitigation within some quadrants.

Both frameworks address the eight risks in the \quadrant{Customer Mandate} quadrant to a certain extent. SAFe appears slightly more
aligned to working with the customer and user than DAD (\cref{fig:mitigation-all}).

\cref{fig:mitigation-all} shows little difference in how the frameworks address the \quadrant{Scope
and Requirements} set of risks.  Both DAD and SAFe address all ten
risks in this quadrant completely, with the exception of DAD that has slightly weaker support for one of the factors, \risk{Users lack understanding of system capabilities and limitations.}

The \quadrant{Execution} quadrant has the most risk factors (31).  As
\cref{fig:mitigation-all} shows, both DAD and SAFe
address most of these 31 risk factors to some extent, although DAD
appears to be the stronger framework when it comes to project execution.

The \quadrant{Environment} quadrant \cref{fig:mitigation-all} has the next highest number of risk factors (since many of the \GSD specific risks are categorized in this quadrant). However, this quadrant has the most factors not addressed by the frameworks. For example, these scaling
agile methods do not appear to fully support \risk{Unstable country/regional political/economic environment}
and \risk{Organization undergoing restructuring during the project.}

When looking across all quadrants far left in \cref{fig:mitigation-all}, the total number of risks we hypothesize each framework addresses are very similar; DAD practices are associated with eliminating (termed ``definitely'') or mitigating (termed ``somewhat'') 58 risks, and SAFe is associated with eliminating (termed ``definitely'') or mitigating (``somewhat'') 57 risks.  We now, in the next section, look to see whether two companies, who implement DAD and SAFe practices, experience any of the associated risks, in order to test our hypotheses.


\subsection{Empirical Evidence }\label{empirical-mapping-results}

To gain some insight into the effectiveness of our theoretical mapping
of scaling agile practices to \GSD risks, we conducted a multiple case study involving two companies engaged in global software development. As described in our method, we examined a range of data collected from these cases for
evidence of how scaling agile practices might have eliminated or
mitigated \GSD risks. 

In this section, we introduce the multiple case study settings and summarize the types of
risk we observed in the cases, as specified in the \GSDRiskCatalog{}.

\subsubsection{Case study setting }\label{sec:case-study-setting}
\begin{figure}
  \centering
  \includegraphics[width=\textwidth]{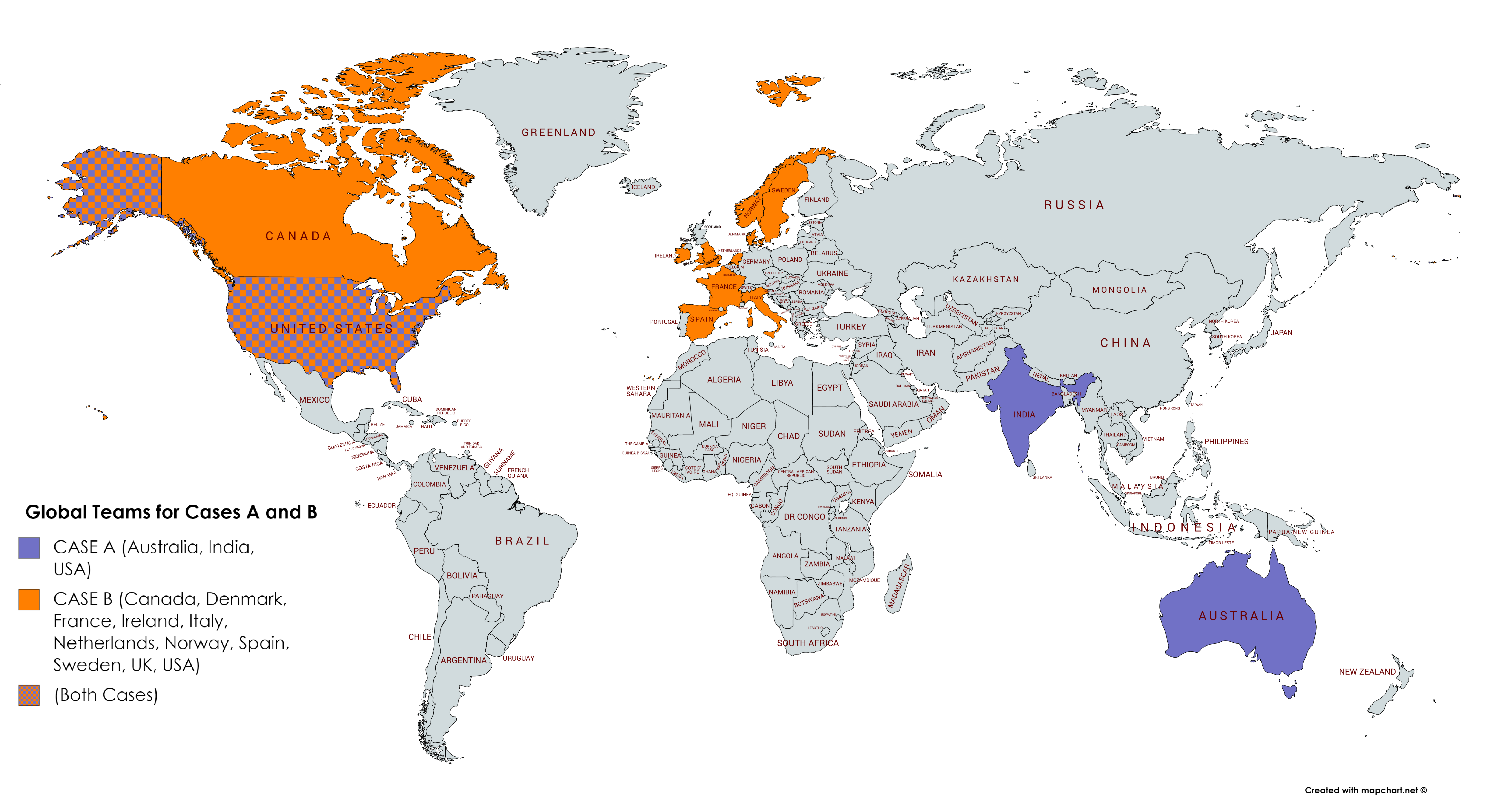}
  \caption{Case locations}\label{fig:case-map}
\end{figure}

Company A, based in Melbourne, Australia, has been using DAD for some
years.  Company B, based in Dublin, Ireland, was undergoing a
transition from a traditional, plan-driven approach to agile
development using SAFe.

As illustrated in \cref{fig:case-map}, both Company A and Company B
have development teams around the world.


\paragraph{Case study A}\label{sec:case-study-a}

Company A is a Melbourne-based company that produces highly intelligent
enterprise asset management software for a global customer base. This
software vendor has undergone a transition towards scaling agile
development using Disciplined Agile Delivery (DAD) over the period from
2015 to 2017, and is continuing as of this writing in 2020. Accompanying this
transition has been a move to provide the software not solely as an
in-house product, but through a cloud-delivered ``software as a
service'' (SaaS) model. The vendor has ten development teams across
three different countries, with the engineering operation based in
Melbourne and development teams in the USA and India (see \cref{fig:case-map}).   The marketing and
after-sales and support teams are based in Australia and the USA.

The company has been actively implementing agile methods since 2003, so
could be considered a mature agile practitioner. Nonetheless, the
transition from their previous ``hybrid-agile'' to a scaling agile
approach using DAD had been driven by pressures of scope and quality:
namely, the inability to deliver the desired scope for planned releases,
and inadequate quality of the software releases delivered into a SaaS
environment.

With their DAD approach, programs run by the Melbourne site involve
their four local project teams 
plus three project teams in the USA, and one in India. Case A is a global software vendor, that meets our
definition of large scale agile \GSD (see \cref{sec:agile-scaling-frameworks}).

The DAD approach has the practice of self-contained teams. So all DAD (project) teams are local and co-located with their own work item lists (product backlogs) allocated by the program management. With the DAD approach a program is delivered through several projects. Hence, the program planning (create program portfolio which basically is the work item list) involving product manager, program manager, and enterprise architects will work at a global level--a program is at a global level whereas projects are local only. While the DAD project teams are local there is a daily tactical huddle where all the leadership roles (Architect, Tech Lead, and Product Owner) of the DAD project teams under a program meet. They (DAD project teams) also do a show and tell collectively for every sprint. Case A is conducting global software development and follows DAD recommendations for self-contained, co-located teams.


\paragraph{Case Study B}\label{sec:case-study-b}

The company we studied for Case Study B is Ocuco Ltd., a medium-sized
Irish software company that develops practice and lab management
software for the optical industry.

Ocuco Ltd., which we will refer to as ``Company B'' in the sequel,
employs approximately 300 staff members in its software development
organization, including support and management staff. Company B has
annual sales exceeding €20 million, from customers in Britain and
Ireland, continental Europe, the Nordic region, North America, and China.

Company B has ten development teams whose members are distributed across
Europe and North America (see~\cref{fig:case-map}), involving approximately 50 developers in
twelve countries; as such, Company B also meets our definition of large
scale agile \GSD (see \cref{sec:agile-scaling-frameworks}).

As part of their transition from a plan-driven
development approach, to agile software development following SAFe,
Company B began introducing Scrum at the team level approximately six
months before we began our study in 2015. SAFe is being rolled out to
the various teams and projects in stages, with the newer projects
leading the way to implementing SAFe practices such as PI Planning,
Automated Testing, and Continuous Integration, whereas much of the
organization is involved in SAFe recommended practices such as
Communities of Practice.

While the purpose of the collaboration with both cases was similar,
which was to observe how teams adopted or transitioned to scaling
agile in a globally distributed setting, there are distinct
differences in the study setting.  Company A has been using agile
methods for nearly two decades, while Company B had just begun a
transition to agile development with SAFe at the time we started our
collaboration.  So, Company A would be considered a mature agile
organization, which at the time of the study was scaling their
development using DAD.  Company B is more of a nascent agile company,
introducing many new agile and lean practices as they attempt to scale
agile development across teams and up the organizational hierarchy.


\subsubsection{Empirical study results}\label{sec:empirical-study-results}


In this section we present results of our investigation into the
extent to which \GSDRiskCatalog risks were observed (or not) in Cases
A and B.  We then establish whether the observed risks in the given
case, were associated with practices  implemented by Company A
or B.  We first provide an overview of this examination showing the
risks not observed, that appear to support our mapping since the risk
may have been eliminated.  Then, we examine the risks that we observed
in the cases.


\begin{singlespace}
\begin{table}[t]
\centering
\caption{Risks from the \GSDRiskCatalog not observed in either Case}\label{tab:risks-not-observed}
\begin{tiny}
\begin{tabular}{p{0.95\textwidth}}
\toprule
\multicolumn{1}{c}{\textbf{Quadrant 1: Customer Mandate}} \tabularnewline
Lack of user participation \tabularnewline
Lack or loss of organizational commitment to the project \tabularnewline
Users not committed to the project \tabularnewline
Users resistant to change \tabularnewline
Users with negative attitudes toward the project \tabularnewline
\multicolumn{1}{c}{\textbf{Quadrant 2: Scope and Requirements}} \tabularnewline
Incorrect system requirements \tabularnewline
Users lack understanding of system capabilities and limitations \tabularnewline
\multicolumn{1}{c}{\textbf{Quadrant 3: Execution}} \tabularnewline
Frequent turnover within the project team \tabularnewline
Ineffective project manager \tabularnewline
Inexperienced project manager \tabularnewline
Inexperienced team members \tabularnewline
Lack of people skills in project leadership \tabularnewline
One of the largest projects attempted by the organization \tabularnewline
\multicolumn{1}{c}{\textbf{Quadrant 4: Environment}} \tabularnewline
Change in organizational management during the project \tabularnewline
Corporate politics with negative effect on project \tabularnewline
Many external suppliers involved in the development project \tabularnewline
new: Unstable country/regional political/economic environment \tabularnewline
Unstable organizational environment \tabularnewline
 
\bottomrule
\end{tabular}
\end{tiny}
\end{table}
\end{singlespace}

\paragraph{Risks not observed in cases as issues}

\begin{table}[t]
\centering
\caption{Summary of empirical evidence supporting theoretical mapping
  of scaling agile practices to risks in the \GSDRiskCatalog.}\label{tab:risks-mapped-impl-seen}
\begin{footnotesize}
\begin{singlespace}
\begin{tabular}{cccrr}
\toprule
Framework  & Practice     & Risk  & \multicolumn{2}{c}{Number of risks} \\
addresses? & implemented? & seen? & Case A & Case B                     \\
\midrule                          
Y          & Y            & N     & 31     & 23                         \\
Y          & N            & Y     & 0      & 0                          \\
Y          & N            & N     & 0      & 0                          \\
Y          & Y            & Y     & 27     & 34                         \\
N          & n/a          & Y     & 0      & 1                          \\
N          & n/a          & N     & 5      & 5                          \\
\bottomrule
\end{tabular}
\end{singlespace}
\end{footnotesize}
\end{table}

Risks shown in \cref{tab:risks-not-observed}, were \emph{not} observed
in the case studies as having become issues; this set of practices
relates to the first row in \cref{tab:risks-mapped-impl-seen}.  This
table provides some evidence
that many risks can be \emph{eliminated} through the adoption of
scaling agile practices. 

This category, where the scaling agile mitigation practice is
implemented, and no related risks are observed, represents the ideal
case, where a risk is identified and addressed before it becomes a
problem.  An example comes from Company A,  involving
spikes\footnote{A `spike' is a ``technical proof-of-concept'' usually
  realized as a throwaway
  prototype~\cite[p. 44]{Ambler_2012_Disciplined}.},  that concerns the risk of
\risk{Team members not familiar with the task(s) being automated}:
``When you got something like what we're doing right now, rolling out
a new dashboard, massive architectural spikes at the start\ldots{} you
got to bring that kind of stuff to the architecture upfront'' (\PA{2}). The
DAD practice \practice{Inception phase involving entire DAD delivery
  team}, along with \practice{Spikes}, ensures that technical
understanding is developed early in the development lifecycle.

Company B employed the SAFe practice \practice{Develop a feature team that
is organized around user-centered functionality} to address a similar
instance of this risk: they employ former users of the product, who have
formal qualifications in the domain, as QA staff; this ensures that
implemented features are usable by actual practitioners.

\begin{figure}
  \centering
  \includegraphics[width=\textwidth]{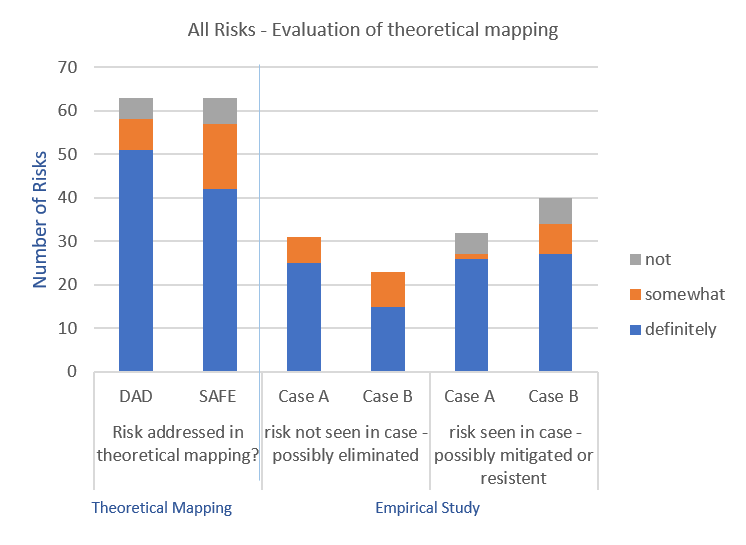}
  \caption{Theoretical mapping evaluation - observations in two cases - All Risks} \label{fig:mitigation-strength-all}
\end{figure}

The proportions of these risks are shown in
\cref{fig:mitigation-strength-all} (middle bars).  In Case A,
practices associated with \numCaseAimpl risks were implemented, yet
less than half (\numCaseAseen) of these risks were observed in Case A;
this suggests that DAD was effective at eliminating over half
(\fracCaseAnotSeen) of the risks for which Company A implemented
associated DAD practices.  Similarly, in Case B, practices associated
with \numCaseBimpl risks were implemented, with \numCaseBnotSeen of
these risks not observed in the case, suggesting that SAFe helped
eliminate up to \fracCaseBnotSeen of risks for which Company B
implemented associated SAFe practices.

\begin{table}
  \centering
  \caption{Risks from \GSDRiskCatalog observed in cases.}\label{tab:risks-observed}
  \begin{singlespace}
  \begin{tiny}
  \begin{tabular}{p{.7\textwidth}cc}
    \toprule
      & \multicolumn{2}{c}{\textbf{Observed in}}\tabularnewline
    \textbf{\GSDRiskCatalog risk} & \textbf{Case A} & \textbf{Case B}\tabularnewline
    \midrule
\multicolumn{3}{c}{\textbf{Quadrant 1: Customer Mandate}} \tabularnewline
Conflict between users & Y &  \tabularnewline
Lack of cooperation from users &  & Y \tabularnewline
Lack of top management support for the project &  & Y \tabularnewline
Lack of user participation &  &  \tabularnewline
Lack or loss of organizational commitment to the project &  &  \tabularnewline
Users not committed to the project &  &  \tabularnewline
Users resistant to change &  &  \tabularnewline
Users with negative attitudes toward the project &  &  \tabularnewline
\multicolumn{3}{c}{\textbf{Quadrant 2: Scope and Requirements}} \tabularnewline
Conflicting system requirements & Y & Y \tabularnewline
Continually changing project scope/objectives &  & Y \tabularnewline
Continually changing system requirements &  & Y \tabularnewline
Difficulty in defining the inputs and outputs of the system & Y &  \tabularnewline
Ill-defined project goals &  & Y \tabularnewline
Incorrect system requirements &  &  \tabularnewline
System requirements not adequately identified & Y & Y \tabularnewline
Unclear system requirements &  & Y \tabularnewline
Undefined project success criteria &  & Y \tabularnewline
Users lack understanding of system capabilities and limitations &  &  \tabularnewline
\multicolumn{3}{c}{\textbf{Quadrant 3: Execution}} \tabularnewline
Development team unfamiliar with selected development tools & Y &  \tabularnewline
Frequent conflicts among development team members &  & Y \tabularnewline
Frequent turnover within the project team &  &  \tabularnewline
High level of technical complexity & Y & Y \tabularnewline
Highly complex task being automated & Y &  \tabularnewline
Immature technology & Y &  \tabularnewline
Inadequate estimation of project budget &  & Y \tabularnewline
Inadequate estimation of project schedule &  & Y \tabularnewline
Inadequate estimation of required resources & Y & Y \tabularnewline
Inadequately trained development team members & Y & Y \tabularnewline
Ineffective communication & Y & Y \tabularnewline
Ineffective project manager &  &  \tabularnewline
Inexperienced project manager &  &  \tabularnewline
Inexperienced team members &  &  \tabularnewline
Lack of an effective project management methodology & Y & Y \tabularnewline
Lack of commitment to the project among development team members &  & Y \tabularnewline
Lack of people skills in project leadership &  &  \tabularnewline
Large number of links to other systems required &  & Y \tabularnewline
Negative attitudes by development team & Y & Y \tabularnewline
new: Ineffective collaboration & Y & Y \tabularnewline
new: Ineffective coordination & Y & Y \tabularnewline
new: Lack of trust & Y & Y \tabularnewline
One of the largest projects attempted by the organization &  &  \tabularnewline
Poor project planning &  & Y \tabularnewline
Project affects a large number of user departments or units & Y &  \tabularnewline
Project involves the use of new technology & Y &  \tabularnewline
Project involves use of technology that has not been used in prior projects &  & Y \tabularnewline
Project milestones not clearly defined &  & Y \tabularnewline
Project progress not monitored closely enough & Y &  \tabularnewline
Team members lack specialized skills required by the project & Y & Y \tabularnewline
Team members not familiar with the task(s) being automated & Y &  \tabularnewline
\multicolumn{3}{c}{\textbf{Quadrant 4: Environment}} \tabularnewline
Change in organizational management during the project &  &  \tabularnewline
Corporate politics with negative effect on project &  &  \tabularnewline
Dependency on outside suppliers &  & Y \tabularnewline
Many external suppliers involved in the development project &  &  \tabularnewline
new: Country-specific regulations & Y & Y \tabularnewline
new: Delays caused by global distance & Y & Y \tabularnewline
new: Lack of architecture-organization alignment & Y & Y \tabularnewline
new: Lack of face-to-face interaction inhibits knowledge sharing & Y & Y \tabularnewline
new: Lack of process alignment & Y & Y \tabularnewline
new: Lack of tool/infrastructure alignment & Y &  \tabularnewline
new: Unstable country/regional political/economic environment &  &  \tabularnewline
Organization undergoing restructuring during the project &  & Y \tabularnewline
Resources shifted from the project due to changes in organizational priorities &  & Y \tabularnewline
Unstable organizational environment &  &  \tabularnewline
 
    \bottomrule
  \end{tabular}
  \end{tiny}
  \end{singlespace}
\end{table}


\paragraph{Risks observed in cases}

\cref{tab:risks-observed} shows that many \GSD risks were observed to
have materialized into issues in one or both cases.  The evidence of
risks being present is drawn from interview transcripts.  

To understand whether the implementation of scaling agile practices
affects the occurrence of risks, we assessed the frequency with which
each company implemented each practice.  Company A has been using DAD
for more than five years; as such, Company A always performs all
except five practices in DAD that address risks in the
\GSDRiskCatalog.  Evidence of practice implementation by Company B
comes from self-assessment surveys.

Company B  was in the middle of a transition to
SAFe when we began our study; consequently, they perform the SAFe
practices at different frequencies, ranging from ``never'' or
``rarely,'' through ``occasionally'' and ``often,'' to ``very often'' or
``always''. We were able to establish the extent to which the SAFe
practices were implemented through a series of self-assessment
surveys~\cite{Razzak_2017_Transition, Razzak_2018_Scaling}
(participants detailed in \cref{tab:list-of-participants-B}).    Based
on these survey responses, we determined that, in Case B six risks had
no associated practices that were performed more often than ``rarely.''

\begin{table}[t]
\centering
\caption{Risks from the \GSDRiskCatalog for which the associated
scaling agile practices were never or rarely performed in our cases.}\label{tab:risks-not-impl}
\begin{singlespace}  
\begin{tiny}
\begin{tabular}{p{0.95\textwidth}}
\toprule
\multicolumn{1}{l}{\emph{Case A}} \tabularnewline
\midrule
\multicolumn{1}{c}{\textbf{Quadrant 3: Execution}} \tabularnewline
Large number of links to other systems required \tabularnewline
\multicolumn{1}{c}{\textbf{Quadrant 4: Environment}} \tabularnewline
Dependency on outside suppliers \tabularnewline
Many external suppliers involved in the development project \tabularnewline
new: Unstable country/regional political/economic environment \tabularnewline
Organization undergoing restructuring during the project \tabularnewline
\midrule
\multicolumn{1}{l}{\emph{Case B}} \tabularnewline
\midrule
\multicolumn{1}{c}{\textbf{Quadrant 3: Execution}} \tabularnewline
Frequent turnover within the project team \tabularnewline
Ineffective project manager \tabularnewline
Inexperienced project manager \tabularnewline
\multicolumn{1}{c}{\textbf{Quadrant 4: Environment}} \tabularnewline
Change in organizational management during the project \tabularnewline
Unstable organizational environment \tabularnewline
Organization undergoing restructuring during the project \tabularnewline
 
\bottomrule
\end{tabular}
\end{tiny}
\end{singlespace}
\end{table}

\cref{tab:risks-not-impl} lists the risks from the \GSDRiskCatalog for
which either company rarely or never performs \emph{any} of the
associated framework practices (see \cref{sec:dad-risk-mapping},
\cref{tab:dad-mapping} and \cref{sec:safe-risk-mapping},
\cref{tab:safe-mapping} for our theoretical mappings).  These results
are not surprising: in the case of Company A, they have a stable
organization and operate in developed countries with a stable
political and economic environment; they also do not engage outside
suppliers or interface with many systems.

Company B likewise has a stable if rapidly growing organization with experienced management, and experiences very low turnover.
So it is perhaps to be expected that they do not implement practices
aimed at reducing the risks associated with these characteristics.


In Case B, there are instances where associated
scaling agile practices were not (yet) fully implemented.  
There are \numCaseBSeenPartialImpl risks in \cref{tab:YYY} where the mode of Company B's
frequency of performance of the associated practices is
``often'' (`3' is ``often'' performed, `4' is ``very often'') indicating they don't always perform these practices.  
As such, risks associated with these practices might not be
fully addressed, and so could be expected to become problems occasionally.
Conversely, Case A always performs associated DAD practices.  This might account
for the fact that fewer risks (\numCaseAseen vs \numCaseBseen) were observed in Case A than
Case B (\cref{fig:mitigation-strength-all}, rightmost bars). 



\cref{tab:risks-mapped-impl-seen} summarizes the results in terms of
the frequencies of combinations of risks addressed and seen, and
practices implemented.  The first column indicates a risk
has been addressed by a scaling agile framework; the second
column indicates whether the mapped scaling agile practices have been
implemented by the company; the third column
indicates a risk has been observed as an issue in the associated case; and, the fourth and fifth columns show the number of
risks that are in the state indicated by the first three columns for
each case.

The first row shows the number of risks from the \GSDRiskCatalog that
have practices mapped from the respective scaling agile framework (DAD
for Case A, and SAFe for Case B), but have not been observed in the respective
case study, and have mapped practices implemented by the case
company.  This row supports the theoretical mapping: the practices
were implemented and the risks were not seen to be present in the case study organization, indicating
that the practices were possibly effective in eliminating the
associated risks.

The second row shows the number of risks that have practices mapped to
them, were observed to have occurred in the respective cases, but the
practices were not implemented.  This row might indicate areas where
the case companies could improve their practices to address observed
risks, but no risks were found in this category.

The third row shows the number of risks that have practices mapped, but were not
observed in the cases, nor were the associated practices implemented.
This category is also empty.
The last two rows show the number of risks that do not have associated
scaling agile practices.  In all but one instance these risks were not
observed in either case.  The one risk observed in this category--from Case B--is \risk{Organization undergoing restructuring during the
project.}  This risk stemmed from the transition from a waterfall to
agile development approach: the product owner of a project focused on
a large customer noted, ``So, we worked in waterfall fashion in
the past and I think this is difficult for people to move from the
waterfall way to the Scrum way'' (\transcriptTwoSCAL).


\begin{table}
\centering
\caption{Framework risk mitigating practice empirical evaluation. ``Degree of impl.'' column indicates
  frequency practice is performed (`3' is ``often'' performed, `4' is ``very often'' 
  performed, and `5' is ``always'' performed).  Table includes risks
  observed in one or both cases (Y); risks in \emph{italics} are seen
  in both cases.  Blank entries in columns indicate risk not
  addressed, or not seen.}\label{tab:YYY}
\begin{tiny}
\begin{singlespace}
\setlength{\tabcolsep}{2pt}
\begin{tabular}{p{.50\textwidth}ccc|ccc}
\toprule
Risk & \multicolumn{3}{c}{DAD \& Case A} & \multicolumn{3}{c}{SAFe \& Case B}                       \\
 & Framework  & Degree   & Risk  & Framework  & Degree   & Risk  \\
 & addresses? & of impl. & seen? & addresses? & of impl. & seen? \\
\midrule
\multicolumn{6}{c}{\textbf{Quadrant 1: Customer Mandate}} \tabularnewline
\midrule
Conflict between users & somewhat &           5 & Y & & & \tabularnewline
Lack of cooperation from users &  &  &  & definitely &           4 &  Y \tabularnewline
Lack of top management support for the project &  &  &  & definitely &           3 &  Y \tabularnewline
\midrule
\multicolumn{6}{c}{\textbf{Quadrant 2: Scope and Requirements}} \tabularnewline
\midrule
\emph{Conflicting system requirements} & definitely &           5 & Y & definitely &           5 & Y \tabularnewline
Continually changing project scope/objectives &  &  &  & definitely &           4 &  Y \tabularnewline
Continually changing system requirements &  &  &  & definitely &           4 &  Y \tabularnewline
Difficulty in defining the inputs and outputs of the system & definitely &           5 & Y & & & \tabularnewline
Ill-defined project goals &  &  &  & definitely &           4 &  Y \tabularnewline
\emph{System requirements not adequately identified} & definitely &           5 & Y & definitely &           5 & Y \tabularnewline
Unclear system requirements &  &  &  & definitely &           5 &  Y \tabularnewline
Undefined project success criteria &  &  &  & definitely &           4 &  Y \tabularnewline
\midrule
\multicolumn{6}{c}{\textbf{Quadrant 3: Execution}} \tabularnewline
\midrule
Development team unfamiliar with selected development tools & definitely &           5 & Y & & & \tabularnewline
Frequent conflicts among development team members &  &  &  & somewhat &           4 &  Y \tabularnewline
\emph{High level of technical complexity} & definitely &           5 & Y & definitely &           4 & Y \tabularnewline
Highly complex task being automated & definitely &           5 & Y & & & \tabularnewline
Immature technology & definitely &           5 & Y & & & \tabularnewline
Inadequate estimation of project budget &  &  &  & definitely &           4 &  Y \tabularnewline
Inadequate estimation of project schedule &  &  &  & definitely &           3 &  Y \tabularnewline
\emph{Inadequate estimation of required resources} & definitely &           5 & Y & definitely &           4 & Y \tabularnewline
\emph{Inadequately trained development team members} & definitely &           5 & Y & somewhat &           4 & Y \tabularnewline
\emph{Ineffective communication} & definitely &           5 & Y & definitely &           5 & Y \tabularnewline
\emph{Lack of an effective project management methodology} & definitely &           5 & Y & definitely &           5 & Y \tabularnewline
Lack of commitment to the project among development team members &  &  &  & definitely &           4 &  Y \tabularnewline
Large number of links to other systems required &  &  &  & definitely &           4 &  Y \tabularnewline
\emph{Negative attitudes by development team} & definitely &           5 & Y & somewhat &           4 & Y \tabularnewline
\emph{new: Ineffective collaboration} & definitely &           5 & Y & definitely &           5 & Y \tabularnewline
\emph{new: Ineffective coordination} & definitely &           5 & Y & definitely &           5 & Y \tabularnewline
\emph{new: Lack of trust} & definitely &           5 & Y & definitely &           5 & Y \tabularnewline
Poor project planning &  &  &  & definitely &           4 &  Y \tabularnewline
Project affects a large number of user departments or units & definitely &           5 & Y & & & \tabularnewline
Project involves the use of new technology & definitely &           5 & Y & & & \tabularnewline
Project involves use of technology that has not been used in prior projects &  &  &  & definitely &           3 &  Y \tabularnewline
Project milestones not clearly defined &  &  &  & definitely &           4 &  Y \tabularnewline
Project progress not monitored closely enough & definitely &           5 & Y & & & \tabularnewline
\emph{Team members lack specialized skills required by the project} & definitely &           5 & Y & somewhat &           3 & Y \tabularnewline
Team members not familiar with the task(s) being automated & definitely &           5 & Y & & & \tabularnewline
\midrule
\multicolumn{6}{c}{\textbf{Quadrant 4: Environment}} \tabularnewline
\midrule
Dependency on outside suppliers &  &  &  & definitely &           4 &  Y \tabularnewline
\emph{new: Country-specific regulations} & definitely &           5 & Y & definitely &           5 & Y \tabularnewline
\emph{new: Delays caused by global distance} & definitely &           5 & Y & somewhat &           5 & Y \tabularnewline
\emph{new: Lack of architecture-organization alignment} & definitely &           5 & Y & somewhat &           5 & Y \tabularnewline
\emph{new: Lack of face-to-face interaction inhibits knowledge sharing} & definitely &           5 & Y & definitely &           5 & Y \tabularnewline
\emph{new: Lack of process alignment} & definitely &           5 & Y & definitely &           3 & Y \tabularnewline
new: Lack of tool/infrastructure alignment & definitely &           5 & Y & & & \tabularnewline
Resources shifted from the project due to changes in organizational priorities &  &  &  & somewhat &           4 &  Y \tabularnewline
 
\bottomrule
\end{tabular}
\end{singlespace}
\end{tiny}
\end{table}

By contrast, the fourth row shows the number of risks that have practices mapped,
were observed in the cases, and the practices were also implemented in
the cases.  \cref{tab:YYY} lists the risks involved in the fourth row
of \cref{tab:risks-mapped-impl-seen}.  This table shows that few (3 of 8) risks
from the \quadrant{Customer Mandate} quadrant were seen in either case
when the associated actions were implemented.  Conversely, the
majority (8 of 10) of \quadrant{Scope and Requirements} risks were seen.
Also, nearly two-thirds (9 of 14) of the risks in the \quadrant{Environment}
quadrant were observed in at least one case, and nearly half (6 of 14) were
seen by both cases.  The majority (24 of 44) of the risks listed in
\cref{tab:YYY} are from the \quadrant{Execution} quadrant.  \cref{sec:discussion}
examines the implications of these risks.

\cref{fig:mitigation-strength-all} 
shows the proportions of these categories in rows one and four of
 \cref{tab:risks-mapped-impl-seen} (risks addressed, practices implemented, and issues not
seen for row one, or seen for row four).
Of note is  the greyed portion of risks at the top of
the stacked bar chart (representing those risks with no associated
practices); these only appear in the category of ``risk seen.''  Conversely,
it is only those risks with associated practices that are \emph{not
  seen}.  The case where the risk is seen, and the practice is
implemented, is harder to interpret; we discuss the
possibilities in the next section (\cref{sec:discussion}).

As \cref{fig:mitigation-strength-all} shows, the majority of
risks that were \emph{seen} as issues in either case are also ``definitely'' addressed
(blue portion of the bars), while a slightly higher proportion of
risks \emph{not} seen in both cases were only ``somewhat'' addressed.
This suggests that the theoretical strength of mitigation or elimination does not
strongly affect whether the risk was observed in a case.


It seems some risks cannot be eliminated, but can be mitigated.

\begin{figure}
  \centering
  \includegraphics[width=\textwidth]{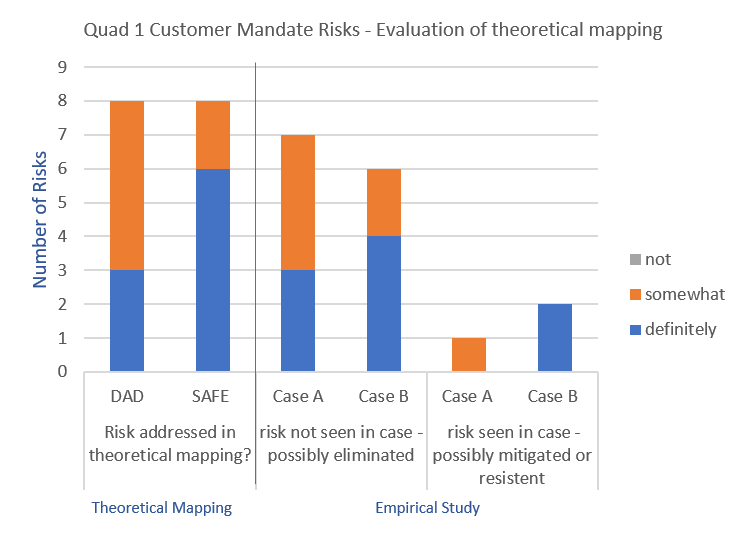}
  \caption{Theoretical mapping evaluation - observations in two cases  - Quad 1} \label{fig:mitigation-strength-q1}
\end{figure}

\cref{fig:mitigation-strength-q1}  shows the proportions of risks
addressed, seen, and associated practices implemented, for the
\quadrant{Customer Mandate} quadrant (quadrant 1).  Only three risks
in this quadrant--\risk{Conflict between users}, \risk{Lack of cooperation from
users}, and \risk{Lack of top management support for the project}--were realized as issues in the cases.

For example, Company A experienced \risk{Conflict between users}: ``Because they've
[product managers] been in the US and I've been here\ldots{} That's sort of things that
people could not appreciate and they wouldn't even hear it because
they know the product, it works like this, and they assumed the
customers wanted the way it was'' (\PA{1}).

An project manager provides an example of  \risk{Lack of cooperation from
users} seen in company B:
 ``\ldots{} now, in an Agile world there is no way that I could tell
 them when they are going to get done until the estimate is there,
 until we start a sprint planning\ldots{} you can't just say, `we are
 doing Agile, so, you got to wait for our next planning\ldots{}'' (\transcriptThreeSCAL).

\begin{figure}
  \centering
  \includegraphics[width=\textwidth]{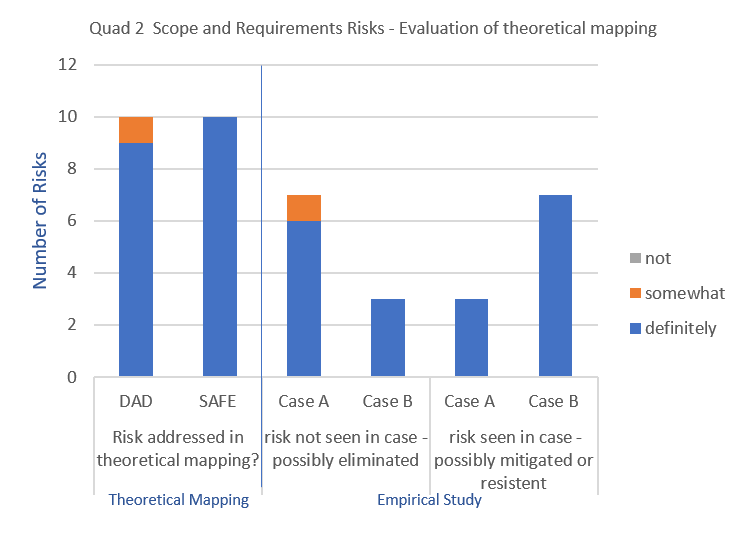}
  \caption{Evaluation of theoretical mapping - observations in two cases - Quad 2} \label{fig:mitigation-strength-q2}
\end{figure}

\cref{fig:mitigation-strength-q2} shows the proportions of risks
addressed, seen, and associated practices implemented, for the
\quadrant{Scope and Requirements} quadrant (quadrant 2).  The
companies in both cases experienced the problems in this quadrant,
including \risk{Conflicting system requirements} and \risk{System
requirements not adequately identified}.

For example, in Case A,
one participant mentioned, ``\ldots{} everything has to be user
stories, sometimes we think that's the underlined problem \ldots{} we
get too focused on that because while it's good to have those stories
to get you going \ldots{} the magic happens every day in the team
making adjustments, embracing that, rather than trying to design user
stories to end.'' (\PA{2})  Similarly, in Case B, a senior manager
noted,
``Like I remember a few weeks ago we were with a customer from
Scotland that asked for a particular improvement in a particular area,
and the whole issue had gotten completely, you know, misunderstood by
development and had been sitting in the backlog for a very long time,
and [we] would set up a call with the customer to try and clear up
what exactly what they want.  Because a lot of the time there is this
`Chinese whispers' thing going on with\ldots{} people misunderstanding
things, recording the wrong request and so on'' (\transcriptFivePMO).

In Case B, the company encountered a problem where the team responsible for
maintaining and enhancing the core product was devoting more effort to
fixing issues raised by customers than to implementing new features;
this caused the core product development to drift away from the product
road-map, resulting in important features being delayed. This is another
example of the  \risk{Conflicting system requirements} risk; the associated SAFe practice,
\practice{Continuously communicate emerging requirements and opportunities
back into the program vision through product owner,} mitigates this risk by ensuring the product vision takes into
account issues raised by customers; also, the product owner would be
fully aware of the product vision and associated road-map, and therefore
would be able to make correct decisions about the relative priorities of
fixes and new features.  As a result of recognizing the link between this
risk and the corresponding SAFe practice, Company B moved from part-time
product owners who doubled as technical support staff, to dedicated
personnel who focus solely on the product owner role.  This does not
\emph{eliminate} the risk of conflicting system requirements, but it
does reduce the impact by ensuring the product owner prioritizes
conflicting requirements properly.

Another example from Case A, relating to conflicting requirements,
shows the power of spikes to mitigate (rather than eliminate) this
risk: an interviewee noted, ``absolutely, it [requirements conflict]
mainly happens in inception, but throughout the project as we get
closer to needing to execute in a specific story, we analyze risks and
potentially do spikes on things, hopefully before those stories are
done'' (\PA{7}).  The DAD practice, \practice{Engineers in the inception phase
  minimize risk through spikes,} helps to resolve conflicts when they
arise.


\begin{figure}
  \centering
  \includegraphics[width=\textwidth]{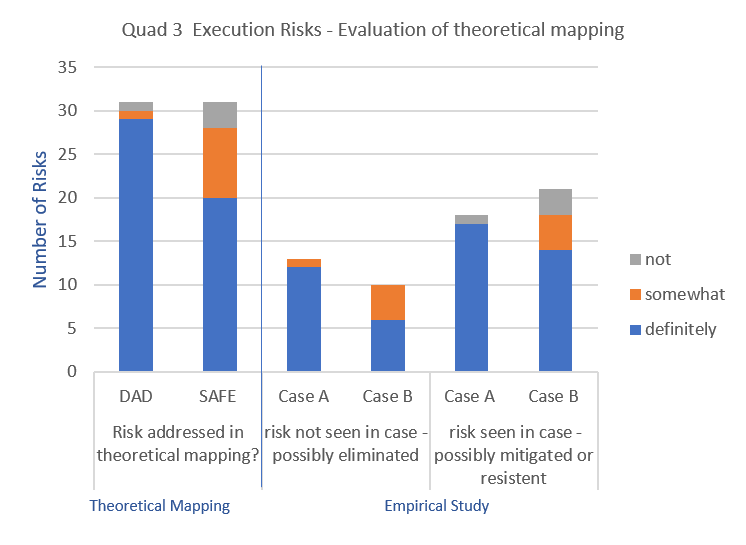}
  \caption{Theoretical mapping evaluation - observations in two cases - Quad 3} \label{fig:mitigation-strength-q3}
\end{figure}

Some risks, all from the \quadrant{Execution} quadrant (quadrant 3, \cref{fig:mitigation-strength-q3}), suggest problems related to agile
development in a \GSD context:
\begin{inparaenum}
\item \risk{Lack of an effective project management methodology}, 
\item \risk{Ineffective coordination},
\item \risk{Ineffective collaboration},
\item \risk{Ineffective communication}, 
\item \risk{Lack of trust},          
\item \risk{Negative attitudes by development team}, 
  and
\item \risk{Inadequate estimation of required  resources}.
\end{inparaenum}

For Case A, \risk{Lack of an effective project management methodology}
manifests itself in technical issues.  For example, one interviewee
described how automation affects their ability to deliver:
``Automation has played a bigger role for us, we reduced the release
cycle from 6 weeks to 2 weeks, without automation to give us that
nightly check ability, it's very difficult, and our product is very
complex, just setting up the environment and do manual testing is
weeks of effort.''  But they are making progress: ``And also, in the
criteria being done, it's as close as possible to 100\%
automation\ldots{} it's something we didn't have before.  We used to
concentrate on a lot of manual work, now we say to the team, it's just
your responsibility as QA to automate'' (\PA{6}).

Both companies experienced
\risk{High level of technical complexity,} and \risk{Team members lack
  specialized skills required by the project,} risks that are related
to technical complexity of development.  For example, a participant
from Case A reflected on the long-term effects of legacy code: ``The
thing that we have to deal with is that we have a lot of legacy code
that was already built in those 3 tiers, and we brought a lot of that
across\ldots{} we were sort of forced to stay with it because the need
to reuse our code, we just didn't have time to build
everything\ldots{} lot of room for improvement I think, if we started
from scratch I think things will be a bit different'' (\PA{2}).  A
scrum master in Case B observed, ``Even though I have [Sr. Developer]
here in Portland [to assist] but he is very heavily involved with
other things.  So, he is quite busy'' (\transcriptTwoSCAL).  Similar
to the \quadrant{Environment} quadrant risks, these two risks are
inherent aspects of the project and so can be mitigated (for example,
by enlisting more experienced developers) but not eliminated.

\begin{figure}
  \centering
  \includegraphics[width=\textwidth]{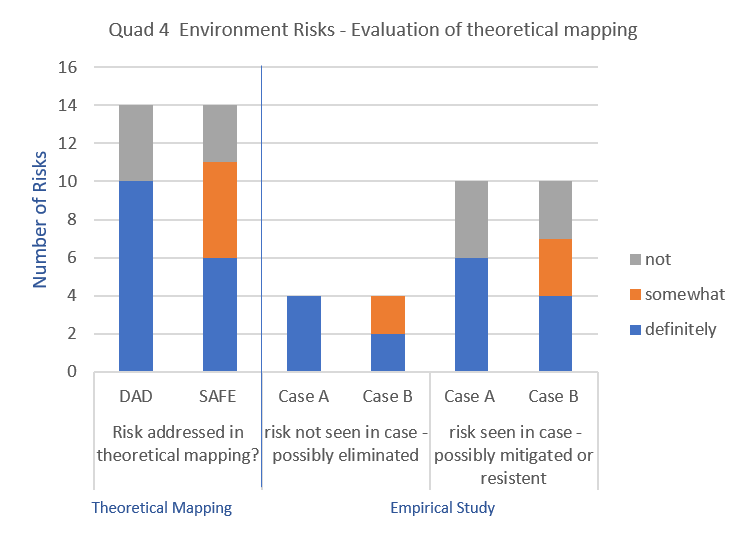}
  \caption{Theoretical mapping evaluation in two cases - Quad 4} \label{fig:mitigation-strength-q4}
\end{figure}

\cref{fig:mitigation-strength-q4} shows the proportions for the
\quadrant{Environment} quadrant (quadrant 4).  
Both companies experienced 
the \risk{Lack of architecture-organization alignment}
risk in this quadrant.  In Case A, this risk manifests itself in their
move to the cloud: ``Yes, we have on the cloud for example, we have
about 5 sort of databases, each hosting a number of tenants, at least
2 different geographies, US and Europe. I think our biggest database
has got like 50 tenants in it, and the other are starting to fill up
as well, and we've reached the limit before, and we just simply start
up another server. It is not a simple matter of one server, you have
to, it's all clustered machines, you have to through 3-4 different
servers you have to bring up to every new database'' (\PA{2}).
In Case B, where growth
by acquisition in different countries has changed the organizational
structure, a developer commented, ``\ldots{} it happens regularly in every week--that someone
forgets to unlock the unit. So, when that happens--previously all
developers were in Dublin and that shouldn't be a problem--now we have
a problem'' (\transcriptSixSCAL).   This introduces \risk{Delays
  caused by global distance} also in quadrant 4:
 ``The difficulty is that, when I [PO] am here on-site I only get
couple of hours in the morning to deal with Dublin stuff.  If I do not
get the things that I need from them even though these two hours I am
pretty much isolated for the day'' (\transcriptTwoSCAL).


We identified several instances where the risks
observed related to the very process of transition to the new
framework (Case B), or in the process of ongoing adaptation and
adopting new practices as the process matured (Case A). For instance,
among the risks observed in each case we noted that \risk{Inadequate
  estimation of project budget} was an issue, and, for Case A, that
DAD was hoped to offer some solutions ``\ldots{} now the DAD process
is coming'' (\PA{4}).  \risk{Inadequately trained development team members}
appeared to be an issue, again with Case A indicating that new DAD
practices and roles involved ``a completely different way of thinking
for us'' (\PA{2}). 
The
risk that \risk{Team members lack specialized skills required by the
  project} was evidenced differently in Case A as team members
``lacking the expertise of integrating all the different products and
offerings under one company product'' (\PA{2}),  and in Case B ``don't fully
understand [the] job'' (\transcriptTwoSCAL), so both new technical demands and new roles
posed challenges.

In the next section, we discuss the implications of these results.

\section{Discussion}\label{sec:discussion}

This study was motivated by the lack of empirical evidence as to the efficacy of scaled agile frameworks in general, and specifically in managing risk in Global Software Development (GSD) settings, where teams are distributed around the world. Our results in \cref{sec:results} provides a promising set of responses to address this gap, as captured by our research question:
\begin{quote}
\RQ{}
\end{quote}

The comprehensive \GSDRiskCatalog{} we derived complements the earlier software risk framework of Wallace \& Keil \cite{Wallace_2004_Software}, by incorporating a further ten new \GSD specific risks, making a total of \numAllRisks risks.  
The new \GSD specific risks are situated within the \quadrant{Environment} and \quadrant{Execution} quadrants of Wallace \& Keil \cite{Wallace_2004_Software}.

\XXXjn{I've taken the subsubsection headings out of what follows; does
  this make sense?}

The theoretical mapping suggests that the two scaling agile frameworks
investigated, DAD and SAFe, could contribute strongly to eliminating
or mitigating risks in our \GSDRiskCatalog{}.  However, the empirical
assessment of issues related to those risks provides a more nuanced
picture of the frameworks and their strengths and limitations.  On the
one hand, nearly half (\numCaseAnotSeen of \numCaseAimpl or
\fracCaseAnotSeen for Case A, and \numCaseBnotSeen of \numCaseBimpl or
\fracCaseBnotSeen for case B) of risks were \emph{not} seen as issues
for the companies when they implemented the associated practices.
This suggests that the respective scaling agile frameworks are
effective at totally \emph{eliminating} a subset of \GSD risks.



\cref{fig:mitigation-strength-all} shows that a majority of risks are
addressed by both DAD and SAFe, both from a theoretical and empirical
point of view.  The vast majority of risks seen in Case A are
addressed by DAD, while a slightly lower proportion of risks seen in
Case B are addressed by SAFe.  Looking at the figure, in both cases it
appears that the \emph{strength} of mitigation has little influence on
whether the company will experience the risk. 



At the time of our study, Company B was in the middle of their
transition from plan-driven to agile development with SAFe.  As such,
for the risks that were seen as issues in Case B, a majority (\FPprint{\numCaseBSeenPartialImpl} of \numCaseBSeen, or \fracCaseBpartialImpl) of associated SAFe practices were performed less than
``always'' (\cref{tab:YYY}, column five).  So this could account for
the somewhat higher number of issues seen in Case B than in Case A.



The remaining risks (that were observed in the cases) are possibly \emph{mitigated} rather than
eliminated.  That is, the risk became a problem, but its \emph{impact} was
reduced by the associated scaling agile practices.  Wallace and Keil
note that companies have low control over the risks in the  \quadrant{Customer Mandate} (quadrant 1) and
\quadrant{Environment} (quadrant 4) quadrants (see~\cref{fig:wk-quadrants});
we would therefore not expect risks in these quadrants to be
eliminated, because they result from external forces out of control of the companies.

\risk{Delays caused by global distance} is an example of an
\quadrant{Environment} risk: if teams are located in Vancouver and
Dublin, or Melbourne and New York, the only way to eliminate delays
caused by lack of timezone overlap is to shift working hours, or close
one location; neither of these is likely to be practical.  Verner and
colleagues also recognize that global
distance~\cite{Noll_2016_Measuring} introduces difficulties for agile
development, noting that ``Lack of synchronous communication in agile
development causes problems'' and ``Collaboration difficulties caused
by geographic distance in agile development may cause
misunderstandings and conflicts''~\cite[p. 64]{Verner_2014_Risks}.
While Verner and colleagues don't offer any mitigation advice for
these risks (unlike other risks associated with agile methods in
\GSD), we assert that SAFe practices such as \practice{Calculate the
  Cost of Delay} and \practice{Manage and optimize the flow of value
  through the program using various tools, such as the Program and
  Value stream Kanbans and information
  radiators}~\cite{Beecham_2020_TR} can reduce the impact of these
risks by highlighting time-critical information.

\XXXjn{This and the following paragraph might be orthogonal to our RQ
  and so maybe should be deleted.} 
Risks in the \quadrant{Customer Mandate} quadrant (quadrant 1) are
determined by the users, who are also largely out of the control of
the development organization.  This would be the case for the
\risk{Conflict between users} and \risk{Lack of cooperation from
  users} risks.  DAD practices such as \practice{Product manager and
  product owner roles, business case, feature funnel, Mandated DOD
  [definition of done], TDD [test-driven development] practices, and
  including manual systems testing before deployment into production
  environment} and \practice{DevOps practices--collaboration between
  Operations and SE (program and project) teams} reduce the impact of
this risk~\cite{Beecham_2020_TR}, but do not eliminate it.  Verner and
colleagues also recognize continuous integration and test-driven
development as ways to mitigate risks associated with agile development
in \GSD~\cite[Table 10, item 2, p. 64]{Verner_2014_Risks}.

\XXXjn{Is this overkill?  Have we made the point sufficiently above?}
Some risks in the \quadrant{Scope and Requirements} quadrant (quadrant
2) also depend on the customer.  \risk{Continually changing system
  requirements} is an example: if the customer decides to change some
of the requirements, for example in response to changing market
conditions, the development teams need to react accordingly or the
customer will not be satisfied.  SAFe practices such as
\practice{Continuously communicate emerging requirements and
  opportunities back into the program vision through product owner,}
\practice{Work with stakeholders to understand the specific business
  targets behind the user-system interaction,} and \practice{Perform
  system demo as near as possible to the end of the iteration} help
ensure that requirements changes are detected and accounted for as
soon as possible.  DAD has similar practices, including having a
\practice{product manager who does market investigation and feedback
  on potential features and functionalities from potential and current
  customers} and a \practice{product owner with UX responsibility, for
  storyboarding, do user research in the field}. 

We note that overall, few risks in the
\quadrant{Customer Mandate} quadrant were seen by either company
(\cref{fig:mitigation-strength-q1}), despite this quadrant being
classified as having a ``low'' perceived level of
control~\cite{Wallace_2004_Software} (\cref{fig:wk-quadrants}).
Conversely, Company B experienced nearly two-thirds (18 of 28) of the
risks in the \quadrant{Scope and Requirements} quadrant (see
\cref{fig:mitigation-strength-q3}), despite this quadrant having a
``high'' perceived level of control~\cite{Wallace_2004_Software}.

Agile methods emphasize the need to \emph{accept} rather than prevent
or control requirements change through requirements freezes and
control change boards in order to deliver the most value to customers
and end-users~\cite{Beck_2000_Extreme}; possibly, this shift in
attitude towards requirements and users means that Wallace and Keil's
original classification of the perceived level of control of the
\quadrant{Customer Mandate} and \quadrant{Scope and Requirements}
quadrants appears to be reversed in the context of agile software
development.




Agile methods also promote intense interactions (both formal and informal) among stakeholders, including the development organization, customers, and users.
This means that there is a closer relationship between developers and
other stakeholders than in traditional plan-driven approaches.
Yet, according to Pikkarainen et al agile methods applied in larger development situations involving multiple external stakeholders ``can sometimes even hinder the communication''~\cite{Pikkarainen_2008_Impact}.

In our theoretical mapping, we contended that many scaling agile practices, such as ``all hands'' PI ceremony, recognize the importance of all distributed team members meeting face-to-face, and that coupled with enhanced tools for video-conferencing and information sharing, and daily remote stand-up meetings, would have contributed to a highly collaborative environment.
Yet the major issues experienced by both cases appear in the \quadrant{Environment} and \quadrant{Execution} quadrants. This may be due to the situation where the ``application of agile practices causes problems in distributed development because of the degree of interaction between stakeholders and number of face-to-face meetings needed''~\cite{Verner_2014_Risks}.  It would appear that \GSDlong impedes this kind of interaction~\cite{Noll_2010_Global}: of
the 17 risks observed in both cases, eight are the new
\GSD risks added to the \GSDRiskCatalog to augment Wallace and Keil's inventory:
\begin{compactenum}
\item \risk{Ineffective collaboration},
\item \risk{Ineffective coordination}, 
\item \risk{Lack of trust},
\item \risk{Country-specific regulations},
\item \risk{Delays caused by global distance},
\item \risk{Lack of architecture-organization alignment},
\item \risk{Lack of face-to-face interaction inhibits knowledge
    sharing}, and
\item \risk{Lack of process alignment}.
\end{compactenum}

\risk{Lack of architecture-organization alignment} is a recurring
issue in
\GSD~\cite{Sievi-Korte_2019_Challenges} that
appears to be a risk that can at best be mitigated; recent studies into
architectural design in \GSD indicate that the architecture does not
always reflect the structure of the organization, where architects
interviewed stated that working across geographic boundaries required
new strategies~\cite{Sievi-Korte_2019_Software}.

Also, both cases experienced \risk{Inadequate estimation of required
  resources.}  Estimation is a persistent problem in any software
development context; in their study of 145 software projects,
Kitchenham and colleagues found that less than two-thirds of the
projects produced estimates within 25\% of the actual time
required~\cite{Kitchenham_2002_Empirical}.  So, inadequate estimation
may be a fact of life in software development, that cannot be
eliminated, especially in a global setting.  

Agile methods accept that
initial estimates are not accurate, but that they
will improve as development teams gain more experience with the
requirements and their own capabilities.  As such, this risk is one
that is likely to be a problem initially, but will be mitigated over time; 
a scrum master in Case B confirmed this, observing, ``For the first
few sprints we completely over-committed to a lot of stuff which we
just couldn't deliver.  So, we are trying to fit in with the velocity
that is based on the size of the team'' (\transcriptTwoSCAL).

It is surprising that \risk{Lack of trust} is an issue in both cases,
since the Agile Manifesto values ``individuals and interactions over
processes and tools,'' and among its twelve principles are ``Build
projects around motivated individuals.  Give them the environment and
support they need, and trust them to get the job
done''~\cite{Beck_2001_Manifesto}.  
As noted in \cref{sec:empirical-study-results}, in
Case A, this
is related to \risk{Ineffective communication} that stems from \GSD,
while in Case B, some of the problem derives from the fact that the company is
still in the process of moving from a plan-driven to an agile
development approach.

Verner and colleagues also recognized that \GSD presents difficulties
for agile software development.
Six risks from Verner et al's
list \cite[Table 10, p. 64]{Verner_2014_Risks} specifically concern agile development; five of these
explicitly cite the impact of \GSD on communication, collaboration, or
coordination: 
\begin{inparaenum}
\item \emph{``Application of agile practices causes problems in distributed
development because of the degree of interaction between
stakeholders and the number of face-to-face meetings needed,''}
\item \emph{``Lack of synchronous communication in agile development
causes problems,''}
\item \emph{``Collaboration difficulties caused by geographic distance in
agile development may cause misunderstandings and
conflicts,''} 
\item \emph{``Poor communication bandwidth for agile development
causes problems with communication and knowledge
management,''}
\item \emph{``Lack of tool support for agile development causes problems
with agile practices,''} and
\item \emph{``Large teams involved with agile development can cause
problems related to communication and coordination.''}
\end{inparaenum}

The fact that \emph{both} cases experienced these risks suggests that
certain risks are endemic to \GSD,
 and, due to their impact on
communication among developers, customers, and users, may be beyond
the capabilities of scaling agile frameworks to \emph{eliminate}.

However, we know from observation that both companies can
exploit communication and coordination technologies such as video
conferencing, real-time ``chat,'' and issue management software, to
effectively implement scaling agile practices related to
communication, coordination, and collaboration.  And we know from
associated studies with Case B,  that practitioners are highly
motivated (with a few exceptions) \cite{Noll_2017_Motivation}, a
status further supported by very low staff turnover
\cite{Bass_2018_Employee, Beecham_2015_What, Hall_2008_Impact}.  
\XXXsb{would be better to note the DAD and SAFe precise practices that
  we know (a) map to the risk from our theoretical mapping, and (b)
  the companies actually implement?  JN: have we not explored this
  exact situation in exhausting detail in the results? sb: Let's leave it. Will mark as resolved} 
So even though we
still observed instances of risks related to communication and
coordination becoming issues, this does not mean that scaling agile
frameworks are ineffective at addressing \GSD risks; rather, we
hypothesize that scaling agile practices can reduce the
\emph{probability} of \GSD risks becoming problems, and potentially
reduce the \emph{impact} of such problems when they do occur, but that
these risks cannot be eliminated entirely.

In summary, the scaling agile philosophy of collaboration, both horizontally and vertically
throughout the enterprise \cite{Leffingwell_2016_Safe, Razzak_2018_Scaling}, appears to eliminate or mitigate software development
risks through better sharing of information, joint decision-making,
progressive refinement, and adaptation of goals.  As noted in Ambler and
Lines~\cite[p. 8]{Ambler_2012_Disciplined}, ``high collaboration is a
hallmark of agility.''  A strong governance structure, and set of
supporting roles and practices within each framework, enable such
behavior, and contribute to reducing risk and thereby to the improved
outcomes advocated by each method~\cite{Noll_2016_Global, Lal_2017_scaling}.


\subsection{Threats to Validity}\label{sec:threats-to-validity}

\subsubsection{Construct Validity}\label{sec:construct-validity}

We have used qualitative data from interviews and observations to
determine whether risks from our \GSDRiskCatalog were experienced by
either of the companies in our case studies. This data collection did
not specifically set out to examine risks that the organizations were
experiencing.  However, the broader questions asked as part of these
longitudinal studies were to understand the challenges and issues that
the practitioners experienced in their development, whilst
transitioning to their target scaling agile frameworks, in a \GSD
setting. 

The measures we used to test whether a scaling agile practice mitigated a given risk, was to look for the absence of qualitative evidence that a company experienced
a risk, combined with evidence that the company had implemented
practices deemed to address the risk, as an indication that the
practices did address the risk by eliminating it. But this relationship may be coincidental;  the risk might not
have materialized anyway, or might have been eliminated by some other means.

Also, we speculated that the \emph{presence} of risks combined with practice
implementation may indicate that the practices do not
eliminate risks.  Again, this could be coincidental.

Future work could involve following-up with the participants to verify
that the scaling agile practices helped, or did not help, to
eliminate risks, and why.

\subsubsection{Internal Validity}\label{sec:internal-validity}

Our approach to mapping agile practices to risks, which we deemed to
mitigate them, was careful, involved inter-rater cross-checks, and was
supported by evidence from each framework's documentation and case data.
But, we may have misinterpreted some of the risks and corresponding
practices.  Moreover, the complexity and interlocking nature of many
roles and practices were such that making a direct mapping was
challenging in some cases. To what extent the identified practices were
mutually self-reinforcing across cases was hard to determine especially since the cases represent different application domains. The observed  differences and similarities therefore, across frameworks, may be due to situational differences rather than framework differences. 

There are trade-offs between types of case study \cite{Bass_2018_Experience} where for example a multiple case study has several validity gains over a single case study in having higher transferability (broader relevance to other cases, and observations are made in a naturalistic real-world setting) and higher confirmability (with more support for findings across studies) \cite{Lincoln_Naturalistic_1985}.  Yet, the level of credibility is considered lower, 
where, for example, confusion can exist in cause and
effect across cases due to the varied settings, that are only visited at one
point in time \cite{Leonard-Barton_1990_Dual}   

As noted above, the presence or absence of risks might not be related to
scaling agile framework practices, but rather could be the result of
confounding variables.  For example, we did not observe the \risk{Unstable
  country/regional political/economic environment} risk in either
case.  But this is almost certainly because both
companies are located in, and sell to, stable economies and countries.
Similarly, Case A experienced the \risk{Project involves the use of
  new technology} risk; since, at the time of our study, they were moving
to SaaS, this risk may simply have been new enough that it was not yet mitigated.

There is also a possibility that researcher bias influenced the 
mappings that form the core of this study.  To reduce this
possibility, we first did independent mappings as individuals, then
compared the results.  We resolved all disagreements by discussion
within each pair of researchers, thus achieving complete agreement in
the end.

The data collected from Case A might be limited by the fact that
interviews and observations were done in one location.  Although the
Melbourne site has interactions with teams and customers across the
globe, and so the participants would have a good understanding of
Company A's processes, they might not have a complete view of the
issues faced by Company A, nor the real frequency with which DAD
practices are performed at other sites.  As this would tend to
underestimate the issues, and overestimate the frequency of
performance, we feel our interpretations  stemming from Case A are
conservative.

\subsubsection{External Validity}\label{sec:external-validity}

Our findings suggest that these conclusions may hold for other companies
implementing scaling agile frameworks, but in particular, the DAD and SAFe
frameworks, as we observed that two companies with different
characteristics and product domains nevertheless experienced similar
risks, especially as related to \GSDlong.
That said, each organization's
implementation will inevitably have its own characteristics, as DAD
and SAFe 
are large, complex, and adaptable frameworks, frequently supported by
experienced agile coaches~\cite[p. 73]{Ambler_2012_Disciplined}, who
are needed to help tailor the frameworks to
the circumstances of each adopting organization. Therefore, the
contributing role of context would need further investigation through
additional studies.

\section{Conclusion}\label{sec:conclusion}

In this paper, through a three-phase process, we have illustrated
how two scaling agile frameworks--DAD and SAFe--largely address the
\numAllRisks software development risks in the \GSDRiskCatalog.

The first phase involved identifying \GSDlong risks faced by software
development organizations, by examining the literature on risks in both
conventional and \GSD contexts.  The result was a \GSDRiskCatalog of 
\numAllRisks risks, divided into four quadrants following Wallace
and Keil~\cite{Wallace_2004_Software}: 
\begin{inparaenum}
\item \quadrant{Customer Mandate},
\item \quadrant{Scope and Requirements},
\item \quadrant{Execution}, and
\item \quadrant{Environment}.
\end{inparaenum}

The next phase consisted of a theoretical mapping of scaling agile practices
to risks in the \GSDRiskCatalog.  We compared practices from DAD and SAFe to risks in the \GSDRiskCatalog, creating a mapping of
practices to risks that shows how scaling agile practices eliminate or
mitigate those risks.

To assess the strength of the scaling agile frameworks to mitigate or
eliminate risk, and avoid criticism that we ``speculated that the
strategy would have helped observed
problems''\cite{Verner_2014_Risks}, we performed an empirical
assessment of the result- ing theoretical mappings.  This empirical
assessment determined the frequency with which practices in each
framework were performed in two companies, and the risks encountered
by those companies.  Through examination of observation and interview
notes and transcripts, and self-assessment survey results, from two
case studies of global software companies, we were able to support
much of our theoretical mapping, and provide evidence that both DAD
and SAFe appear to eliminate or mitigate the majority of risks in the
\GSDRiskCatalog.

Thus, this study adds to the limited empirical evidence of the
efficacy of scaling agile frameworks.  It suggests that  the claims
that these frameworks are risk and value-driven approaches have
some validity.

Of the four quadrants in the \GSDRiskCatalog, the \quadrant{Customer
Mandate} quadrant appears to be better addressed through the SAFe framework than
DAD. \quadrant{Scope and Requirements} risks are addressed well by both
methods. \quadrant{Execution} risks are better mitigated by DAD than SAFe,
and \quadrant{Environment} risks are less well addressed by either
approach.  This suggests that the \quadrant{Environment} set of risks
are less amenable to being addressed by a process framework.

A further outcome of creating the \GSDRiskCatalog is the addition of
ten new risks related to \GSDlong, which were not identified in the
Wallace and Keil inventory~\cite{Wallace_2004_Software} (see
\cref{tab:new-risks}).  These new risks appear to be endemic and suggest a \emph{risk tariff} in \GSD;
all of these except  \risk{Lack of tool/infrastructure alignment} and \risk{Unstable
  country/regional political/economic environment} were experienced by
\emph{both} companies. 

The result of these three phases is a scaling agile risk theoretical mapping that
shows how two scaling agile frameworks--Disciplined Agile Delivery and
the Scaled Agile Framework--can potentially eliminate or mitigate
software project risks in global software
development.  

These findings in a global software development scaling agile context
echo Oehmen and colleagues'~\cite{Oehmen_2014_Analysis} assertion that
risks cannot be avoided; at best organizations manage risk by applying
practices that lead to a structured reduction of uncertainty.

\subparagraph{Acknowledgment}\label{sec:acknowledgment}

We are indebted to our Case Study organizations--Company A and Ocuco Ltd (Company B)--and the
many members of those organizations we observed and interviewed over a
four-year period. Extra special thanks to Clodagh Nic Canna, of Ocuco, who was the main driver behind our collaboration. We also offer our thanks to Abdur Razzak, who helped
with data collection and some early analysis of issues.
This work was supported, in part, by Science Foundation Ireland grant
13/RC/2094.

\bibliographystyle{elsarticle-num}
\bibliography{ms}

\appendix
\def\appendixname{}             

\section{DAD Practice Mapping to \GSDRiskCatalog risks}\label{sec:dad-risk-mapping}

\begin{tiny}
\begin{singlespace}

\begin{longtable}[t]{p{.75\textwidth}l}
  \caption{DAD practices mapped to \GSDRiskCatalog risks}\label{tab:dad-mapping}\tabularnewline
  \toprule
  \emph{\GSDRiskCatalog risk}/Practices & \emph{Level}\tabularnewline
  \endfirsthead
  \toprule
  \emph{\GSDRiskCatalog risk}/Practices & \emph{Level}\tabularnewline
  \midrule
  \endhead
\midrule
\multicolumn{2}{c}{\textbf{Quadrant 1: Customer Mandate}} \tabularnewline
\midrule
\emph{Conflict between users} & somewhat \tabularnewline
\midrule
\multicolumn{2}{p{\textwidth}}{Product manager responsible for features and functionalities, product owner with UX responsibility.  Business case, feature funnel, storyboarding, user stories, DOD including acceptance tests, iteration show and tell, beta testing, production environment test, DevOp practices, real time monitoring of the product environment.  } \tabularnewline
\midrule
\emph{Lack of cooperation from users} & somewhat \tabularnewline
\midrule
\multicolumn{2}{p{\textwidth}}{Product manager, product owner.  Business case, feature funnel, storyboarding, DevOp practices, real time monitoring of the product environment by end-users, feedback from end-users.  } \tabularnewline
\midrule
\emph{Lack of top management support for the project} & definitely \tabularnewline
\midrule
\multicolumn{2}{p{\textwidth}}{Portfolio manager, portfolio planning team. Portfolio planning including feature funnel practices, business case including six monthly presentation to the executives for approval.  } \tabularnewline
\midrule
\emph{Lack of user participation} & somewhat \tabularnewline
\midrule
\multicolumn{2}{p{\textwidth}}{Product management, vision planning, program management, story boarding.  Product manager, proxy user product owner, on-site customer with UX responsibility.  } \tabularnewline
\midrule
\emph{Lack or loss of organizational commitment to the project} & definitely \tabularnewline
\midrule
\multicolumn{2}{p{\textwidth}}{Product management, portfolio management, program management teams.  Product planning, portfolio planning, and program planning.  Work item list.  } \tabularnewline
\midrule
\emph{Users not committed to the project} & somewhat \tabularnewline
\midrule
\multicolumn{2}{p{\textwidth}}{Product manager, product owner, domain expert.  Business case, feature funnel, storyboarding, DevOp practices, real time monitoring of the product environment by end-users, feedback from end-users.  } \tabularnewline
\midrule
\emph{Users resistant to change} & definitely \tabularnewline
\midrule
\multicolumn{2}{p{\textwidth}}{Product management, vision planning,  program management, story boarding.  Product manager, proxy user product owner, on-site customer with UX responsibility.   Both roles require individuals with solid local technical product knowledge.  } \tabularnewline
\midrule
\emph{Users with negative attitudes toward the project} & somewhat \tabularnewline
\midrule
\multicolumn{2}{p{\textwidth}}{Product manager of features and functionalities, product owner with UX responsibility).  Business case, feature funnel, storyboarding, user stories, DOD including acceptance tests, iteration show and tell, beta testing, production environment test, DevOp practices, real time monitoring of the product environment.  } \tabularnewline
\midrule
\multicolumn{2}{c}{\textbf{Quadrant 2: Scope and Requirements}} \tabularnewline
\midrule
\emph{Conflicting system requirements} & definitely \tabularnewline
\midrule
\multicolumn{2}{p{\textwidth}}{Product manager who does market investigation and feedback on potential features and functionalities from potential and current customers.  Vision planning, product owner with UX responsibility, for storyboarding, do user research in the field.  Engineers in the inception phase  minimise risk through spikes. Vision planning, business case, feature funnel.  } \tabularnewline
\midrule
\emph{Continually changing project scope/objectives} & definitely \tabularnewline
\midrule
\multicolumn{2}{p{\textwidth}}{Program management creates work item list for the entire program.  Delivery teams choose user stories for their work item list (project).  Program daily huddles.  DOD, UAT. Programs are time-boxed, defined, and relatively short duration.  } \tabularnewline
\midrule
\emph{Continually changing system requirements} & definitely \tabularnewline
\midrule
\multicolumn{2}{p{\textwidth}}{Product manager who does market investigation and feedback on potential features and functionalities from potential and current customers.  Vision planning, product owner with UX responsibility, for storyboarding, do user research in the field.  Engineers in the inception phase  minimise risk through spikes. Vision planning, business case, feature funnel.  } \tabularnewline
\midrule
\emph{Difficulty in defining the inputs and outputs of the system} & definitely \tabularnewline
\midrule
\multicolumn{2}{p{\textwidth}}{Program management including Enterprise architects and Product owners. Program work item list, architecture design, story boarding, user stories.   Project-level product owners, team architects, software, and QA engineers.  Inception phase including spikes to  understand users stories.  UAT during onstruction phase.  } \tabularnewline
\midrule
\emph{Ill-defined project goals} & definitely \tabularnewline
\midrule
\multicolumn{2}{p{\textwidth}}{Product owner, Team lead, Architecture owner.  Project work item list, DOD, inception phase, construction phase, transition phase.  } \tabularnewline
\midrule
\emph{Incorrect system requirements} & definitely \tabularnewline
\midrule
\multicolumn{2}{p{\textwidth}}{Product management.  Vision planning to identify high level/epic stories.  Product owner,  story boarding.  } \tabularnewline
\midrule
\emph{System requirements not adequately identified} & definitely \tabularnewline
\midrule
\multicolumn{2}{p{\textwidth}}{Program management including Product owners, enterprise architects, program manager.  Program work item list, story boarding, user stories.   Project level product owners, team architects, software and QA engineers.  Inception phase project item list, spikes, user stories estimated and prioritised using user story points,  } \tabularnewline
\midrule
\emph{Unclear system requirements} & definitely \tabularnewline
\midrule
\multicolumn{2}{p{\textwidth}}{Program management including enterprise architects. Program work item list.  Architect owner.  Spikes.  } \tabularnewline
\midrule
\emph{Undefined project success criteria} & definitely \tabularnewline
\midrule
\multicolumn{2}{p{\textwidth}}{Portfolio, program management, project management.  Portfolio manager, program manager, product owners and enterprise architects. Release plan (roadmap) at portfolio level; program level work item.  Iteration show and tell, program daily hurdles, project level DOD and story points.  } \tabularnewline
\midrule
\emph{Users lack understanding of system capabilities and limitations} & somewhat \tabularnewline
\midrule
\multicolumn{2}{p{\textwidth}}{Product managers, customer success team.  Vision planning.  Program management does storyboarding for UX.  } \tabularnewline
\midrule
\multicolumn{2}{c}{\textbf{Quadrant 3: Execution}} \tabularnewline
\midrule
\emph{Development team unfamiliar with selected development tools} & definitely \tabularnewline
\midrule
\multicolumn{2}{p{\textwidth}}{DAD delivery teams are empowered to make tool decision  } \tabularnewline
\midrule
\emph{Frequent conflicts among development team members} & definitely \tabularnewline
\midrule
\multicolumn{2}{p{\textwidth}}{Program management, including program manager, product owners, Enterprise architect, team leads, Architect owners.  Daily tactical huddle.  DAD delivery team.  Primary and secondary roles. Daily stand-up meeting, self-organising teams, empowerment.  } \tabularnewline
\midrule
\emph{Frequent turnover within the project team} & definitely \tabularnewline
\midrule
\multicolumn{2}{p{\textwidth}}{Self-contained DAD delivery team.  Primary and secondary roles, including three leadership roles and separate HR manager.  Self-organising teams, empowerment, task sharing, T-skilling.  } \tabularnewline
\midrule
\emph{High level of technical complexity} & definitely \tabularnewline
\midrule
\multicolumn{2}{p{\textwidth}}{Entire DAD delivery team involved in inception phase. Spikes.  Provide story points and estimates.  Sprint planning.  } \tabularnewline
\midrule
\emph{Highly complex task being automated} & definitely \tabularnewline
\midrule
\multicolumn{2}{p{\textwidth}}{Entire DAD delivery team involved in inception phase. Spikes.  Provide story points and estimates.  Sprint planning.  } \tabularnewline
\midrule
\emph{Immature technology} & definitely \tabularnewline
\midrule
\multicolumn{2}{p{\textwidth}}{DAD delivery teams are empowered to make tool decision.  Temporary roles in DAD delivery teams.  Coaching and training for upskilling on unfamiliar technology.  } \tabularnewline
\midrule
\emph{Inadequate estimation of project budget} & somewhat \tabularnewline
\midrule
\multicolumn{2}{p{\textwidth}}{Features are delivered through program.  } \tabularnewline
\midrule
\emph{Inadequate estimation of project schedule} & definitely \tabularnewline
\midrule
\multicolumn{2}{p{\textwidth}}{User stories and estimates.  Priority setting.  Spikes.  User stories, story points.  Re-estimating.  } \tabularnewline
\midrule
\emph{Inadequate estimation of required resources} & definitely \tabularnewline
\midrule
\multicolumn{2}{p{\textwidth}}{Self contained DAD delivery team with up to 13 individuals.  Primary and secondary roles three leadership roles and separate HR manager.  Empowered self-organising teams.  Task sharing.  T-skilling.  } \tabularnewline
\midrule
\emph{Inadequately trained development team members} & definitely \tabularnewline
\midrule
\multicolumn{2}{p{\textwidth}}{Project management including Product owner, Architecture owner.  Self-organising teams, T-skilled, coaching by Architecture owner and Product owner, pair programming.  Bring in outside expertise to develop and upskill team members.  } \tabularnewline
\midrule
\emph{Ineffective communication} & definitely \tabularnewline
\midrule
\multicolumn{2}{p{\textwidth}}{Project management: empowerment, self contained delivery team  (PO, AO, TL, SE, QA, T. Writer etc), self-organising, pair programming.  } \tabularnewline
\midrule
\emph{Ineffective project manager} & definitely \tabularnewline
\midrule
\multicolumn{2}{p{\textwidth}}{Shared leadership roles (Product owner, Architecture owner, Technical lead) for DAD delivery teams.  Project management tasks are shared by the entire DAD delivery team (individuals in primary roles).  } \tabularnewline
\midrule
\emph{Inexperienced project manager} & definitely \tabularnewline
\midrule
\multicolumn{2}{p{\textwidth}}{Shared leadership roles (Product owner, Architecture owner, Technical lead) for DAD delivery teams.  Project management tasks are shared by the entire DAD delivery team (individuals in primary roles).  } \tabularnewline
\midrule
\emph{Inexperienced team members} & definitely \tabularnewline
\midrule
\multicolumn{2}{p{\textwidth}}{Self-contained DAD delivery team with up 13 individuals.  Primary and secondary roles.  Three leadership roles.  Task sharing, pair programming.  } \tabularnewline
\midrule
\emph{Lack of an effective project management methodology} & definitely \tabularnewline
\midrule
\multicolumn{2}{p{\textwidth}}{Clearly defined program management roles and practices.  Clearly defined project management practices and roles, including DAD method phases and activities.  } \tabularnewline
\midrule
\emph{Lack of commitment to the project among development team members} & definitely \tabularnewline
\midrule
\multicolumn{2}{p{\textwidth}}{Project management including Team lead.  Self-organising teams, inception phase engineers empowered to run spikes,  provide story points and re-estimate each user story.  Sprint planning and committment to sprint objectives.  } \tabularnewline
\midrule
\emph{Lack of people skills in project leadership} & definitely \tabularnewline
\midrule
\multicolumn{2}{p{\textwidth}}{Clearly defined leadership roles and practices with program and project management including based on self-organising DAD delivery teams.  } \tabularnewline
\midrule
\emph{Large number of links to other systems required} & not at all \tabularnewline
\midrule
\emph{Negative attitudes by development team} & definitely \tabularnewline
\midrule
\multicolumn{2}{p{\textwidth}}{Self-organising teams.  Inception phase engineers empowered to run spikes, provide story points, and re-estimate each user story.  Sprint planning and committment to sprint objectives.  } \tabularnewline
\midrule
\emph{new: Ineffective collaboration} & definitely \tabularnewline
\midrule
\multicolumn{2}{p{\textwidth}}{All DAD project teams are  feature team, self-contained and self-organizing teams, T-skilled and all team members are in  primary roles -every one are treated as equal, empowered, and make collective decisions. Three leadership roles with a DAD project team- team lead, AO and PO.  All DAD project teams are  feature team, self-contained and self-organizing teams, T-skilled and all team members are in  primary roles (software engineers). Three leadership roles with a DAD project team- team lead, AO and PO. Co-location, shared workspace, daily stand-up meetings, pair programming.  All DAD project teams are  feature team, self-contained and self-organizing teams, T-skilled and all team members are in  primary roles. Three leadership roles with a DAD project team- team lead, AO and PO. Co-location, shared workspace, daily stand-up meetings, pair programming.  } \tabularnewline
\midrule
\emph{new: Ineffective coordination} & definitely \tabularnewline
\midrule
\multicolumn{2}{p{\textwidth}}{All DAD project teams are  feature team, self-contained and self-organizing teams, T-skilled and all team members are in  primary roles (software engineers). Three leadership roles with a DAD project team- team lead, AO and PO. Co-location, shared workspace, daily stand-up meetings,  All DAD project teams are  feature team, self-contained and self-organizing teams, T-skilled and all team members are in  primary roles (software engineers). Three leadership roles with a DAD project team- team lead, AO and PO. Co-location, shared workspace, daily stand-up meetings, iteration planning, iteration reviews.  All DAD project teams are  feature team, self-contained and self-organizing teams, T-skilled and all team members are in  primary roles (software engineers). Three leadership roles with a DAD project team- team lead, AO and PO. Co-location, shared workspace, daily stand-up meetings, pair programming.  } \tabularnewline
\midrule
\emph{new: Lack of trust} & definitely \tabularnewline
\midrule
\multicolumn{2}{p{\textwidth}}{All DAD project teams are  feature team, self-contained and self-organizing teams, T-skilled and all team members are in  primary roles -every one are treated as equal, empowered, and make collective decisions. Any training need for the team is allowed through the secondary role- a coach, consultant or skilled individual member from another team can join for a period of time  to upskill those who need upskilling.   All DAD project teams are  feature team, self-contained and self-organizing teams, T-skilled and all team members are in  primary roles -every one are treated as equal, empowered, and make collective decisions. T-skilled means every team must learn on the fly all the broad skills required to deliver projects.  All DAD project teams are feature team, self-contained and  self-organizing teams, T-skilled and all team members are in  primary roles only-every one are treated as equal, empowered, and make collective decisions. However, AO, PO and Team lead can make decisions so that teams can make progress if stuck with a problem.  } \tabularnewline
\midrule
\emph{One of the largest projects attempted by the organization} & definitely \tabularnewline
\midrule
\multicolumn{2}{p{\textwidth}}{Program management, including Enterprise architect and product owners.  Work item list and DAD delivery teams.  Iteration show and tell.  } \tabularnewline
\midrule
\emph{Poor project planning} & definitely \tabularnewline
\midrule
\multicolumn{2}{p{\textwidth}}{Program Manager, Enterprise architects, Product owners.  Work item list.  Story boarding.  Architecture planning.  Product owners, Architect owners, Team lead, DAD delivery team.  Inception phase, iteration planning, DOD, UAT.  Re-estimation, story points.  } \tabularnewline
\midrule
\emph{Project affects a large number of user departments or units} & definitely \tabularnewline
\midrule
\multicolumn{2}{p{\textwidth}}{Product management, portfilio management, project management, IT operations including cloud deployment.  } \tabularnewline
\midrule
\emph{Project involves the use of new technology} & definitely \tabularnewline
\midrule
\multicolumn{2}{p{\textwidth}}{DAD delivery teams are empowered to make tool decision.  Technical experts as temporary roles in DAD delivery teams.  Coaching and training for upskilling on unfamiliar technology.  } \tabularnewline
\midrule
\emph{Project involves use of technology that has not been used in prior projects} & definitely \tabularnewline
\midrule
\multicolumn{2}{p{\textwidth}}{DAD delivery teams are empowered to make tool decisions. Temporary roles in DAD delivery teams.  Coaching and training for upskilling on unfamiliar technology.  } \tabularnewline
\midrule
\emph{Project milestones not clearly defined} & definitely \tabularnewline
\midrule
\multicolumn{2}{p{\textwidth}}{Project management- each phase can be a milestone including creating a work item list, delivering  sprint/iteration based on DOD.  } \tabularnewline
\midrule
\emph{Project progress not monitored closely enough} & definitely \tabularnewline
\midrule
\multicolumn{2}{p{\textwidth}}{Program manager responsible to deliver a program.  Program daily huddles.  Iteration show and tell.  Product owner, Architecture owner, team leader.  User story points.  Time-boxed construction phase has several iteration cycles, all time boxed.  Daily stand-up meeting.  } \tabularnewline
\midrule
\emph{Team members lack specialized skills required by the project} & definitely \tabularnewline
\midrule
\multicolumn{2}{p{\textwidth}}{Product owner, Architecture owner.  Self-organising teams, T-skilling, coaching by Architecture owner and Product owner, pair programming.   Bring in outside expertise to develop and upskill team members.  } \tabularnewline
\midrule
\emph{Team members not familiar with the task(s) being automated} & definitely \tabularnewline
\midrule
\multicolumn{2}{p{\textwidth}}{Inception phase involving entire DAD delivery team. Spikes.  Story points and estimates, sprint planning.  } \tabularnewline
\midrule
\multicolumn{2}{c}{\textbf{Quadrant 4: Environment}} \tabularnewline
\midrule
\emph{Change in organizational management during the project} & definitely \tabularnewline
\midrule
\multicolumn{2}{p{\textwidth}}{IT governances drives DAD organisation structures.  Product management, portfolio management,  program management, and project management.  } \tabularnewline
\midrule
\emph{Corporate politics with negative effect on project} & definitely \tabularnewline
\midrule
\multicolumn{2}{p{\textwidth}}{Product management, portfolio management,  program management and project management.  Collective decision making at various levels with DAD framework.  } \tabularnewline
\midrule
\emph{Dependency on outside suppliers} & not at all \tabularnewline
\midrule
\emph{Many external suppliers involved in the development project} & not at all \tabularnewline
\midrule
\emph{new: Country-specific regulations} & definitely \tabularnewline
\midrule
\multicolumn{2}{p{\textwidth}}{DevOp practices- collaboration between Operations and SE (program and project)teams.  Portfolio planning at enterprise level, product management, program planning at engineering level, PO, enterprise architects.  } \tabularnewline
\midrule
\emph{new: Delays caused by global distance} & definitely \tabularnewline
\midrule
\multicolumn{2}{p{\textwidth}}{DevOp practices- collaboration between Operations and SE (program and project)teams.  Mandated DOD, TDD practices and including manual systems testing before deployment into production environment.  Program management- program manager, enterprise architect, PO.  } \tabularnewline
\midrule
\emph{new: Lack of architecture-organization alignment} & definitely \tabularnewline
\midrule
\multicolumn{2}{p{\textwidth}}{DevOp practices at program and project levels. Performance monitoring practice of product environment for real time data for program portfolio planning and project planning.  Enterprise architectures provide technology roadmap and guidance for project teams. Teams must run spikes.  Enterprise architecture team (EAT) [VP architect + 8 enterprise architects (EA) from different geographic locations] plan and provide  architecture strategy and guidance for the entire organization through weekly virtual EAT meetings. provide are Program architects owners form the enterprise architecture team"  } \tabularnewline
\midrule
\emph{new: Lack of face-to-face interaction inhibits knowledge sharing} & definitely \tabularnewline
\midrule
\multicolumn{2}{p{\textwidth}}{All DAD project teams are  feature team, self-contained and self-organizing teams, T-skilled and all team members are in  primary roles (software engineers). Three leadership roles with a DAD project team- team lead, AO and PO. Co-location, shared workspace, daily stand-up meetings, iteration planning, iteration reviews.  DevOp practices- collaboration between Operations and SE (program and project)teams.  Project level (DAD project teams)- inception phase-spikes, story-points and re-estimate, construction phase- look ahead planning, iteration planning- re-estimate  } \tabularnewline
\midrule
\emph{new: Lack of process alignment} & definitely \tabularnewline
\midrule
\multicolumn{2}{p{\textwidth}}{All DAD project teams are  feature team, self-contained and self-organizing teams, T-skilled and all team members are in  primary roles (software engineers). Three leadership roles with a DAD project team- team lead, AO and PO. Co-location, shared workspace, daily stand-up meetings, iteration planning, iteration reviews. Flexibility and empowerment for project teams to also adopt their team specific practices.  DAD project teams- a coach or consultant in secondary role  Product management and program management- mandated practices for projects.  } \tabularnewline
\midrule
\emph{new: Lack of tool/infrastructure alignment} & definitely \tabularnewline
\midrule
\multicolumn{2}{p{\textwidth}}{Booking shared resources.  Mandated DevOp practice, collaboration between operations and engineering teams.  Mandated practice- DevOp practice, collaboration between operations and engineering teams.  } \tabularnewline
\midrule
\emph{new: Unstable country/regional political/economic environment} & not at all \tabularnewline
\midrule
\emph{Organization undergoing restructuring during the project} & not at all \tabularnewline
\midrule
\emph{Resources shifted from the project due to changes in organizational priorities} & definitely \tabularnewline
\midrule
\multicolumn{2}{p{\textwidth}}{Program level defines projects and resources.  DAD delivery teams remain fixed based on small, medium and large teams (each has a defined number of individuals).  } \tabularnewline
\midrule
\emph{Unstable organizational environment} & definitely \tabularnewline
\midrule
\multicolumn{2}{p{\textwidth}}{IT governances drives DAD organisation structures.  Product management, portfolio management,  program management, and project management.  } \tabularnewline
 
  \bottomrule
\end{longtable}
\end{singlespace}
\end{tiny}

\section{SAFe Practice Mapping to \GSDRiskCatalog risks}\label{sec:safe-risk-mapping}

\begin{tiny}
\begin{singlespace}
\begin{longtable}[t]{p{.75\textwidth}l}
  \caption{SAFe practices mapped to \GSDRiskCatalog risks}\label{tab:safe-mapping}\tabularnewline
  \toprule
  \emph{\GSDRiskCatalog risk}/Practices & \emph{Level}\tabularnewline
  \endfirsthead
  \toprule
  \emph{\GSDRiskCatalog risk}/Practices & \emph{Level}\tabularnewline
  \midrule
  \endhead
\midrule
\multicolumn{2}{c}{\textbf{Quadrant 1: Customer Mandate}} \tabularnewline
\midrule
\emph{Conflict between users} & somewhat \tabularnewline
\midrule
\multicolumn{2}{p{\textwidth}}{Help to derive collaboration among the key stakeholders in the business.  Use capabilities as end-to-end solution services that support the achievement of user goals.  Complete user stories from Team Backlog and assure that each story meets its (local) Definition of Done.  } \tabularnewline
\midrule
\emph{Lack of cooperation from users} & definitely \tabularnewline
\midrule
\multicolumn{2}{p{\textwidth}}{Encourage and facilitate incremental development and fast customer feedback.  Facilitate continuous improvement by quantitative metrics, customer feedback, and the Inspect and Adapt retrospective cycle.  Engage stakeholders and provide feedback--time-boxed to an hour or less.  } \tabularnewline
\midrule
\emph{Lack of top management support for the project} & definitely \tabularnewline
\midrule
\multicolumn{2}{p{\textwidth}}{Allocate budget for development of value stream.   Define Strategic Themes.  Allocate budget authority to the decision makers.  } \tabularnewline
\midrule
\emph{Lack of user participation} & definitely \tabularnewline
\midrule
\multicolumn{2}{p{\textwidth}}{Facilitate continuous improvement by quantitative metrics, customer feedback, and the Inspect and Adapt retrospective cycle.  Develop a feature team that is organized around user-centered functionality. Each team is capable of delivering end-to-end value. Feature teams operate primarily with user stories, refactors, and spikes.  Fast and intimate feedback from customers.  } \tabularnewline
\midrule
\emph{Lack or loss of organizational commitment to the project} & definitely \tabularnewline
\midrule
\multicolumn{2}{p{\textwidth}}{Allocate budget for development of value stream.   Define Strategic Themes.  Allocate budget authority to the decision makers.  } \tabularnewline
\midrule
\emph{Users not committed to the project} & definitely \tabularnewline
\midrule
\multicolumn{2}{p{\textwidth}}{Encourage and facilitate incremental development and fast customer feedback.  Facilitate continuous improvement by quantitative metrics, customer feedback, and the Inspect and Adapt retrospective cycle.  Engage stakeholders and provide feedback--time-boxed to an hour or less.  } \tabularnewline
\midrule
\emph{Users resistant to change} & somewhat \tabularnewline
\midrule
\multicolumn{2}{p{\textwidth}}{Provide the opportunity for 'just the right amount' of Architecture and User Experience guidance.  Identify user-business value.  Continuously validate user experience via user experience testing.  } \tabularnewline
\midrule
\emph{Users with negative attitudes toward the project} & definitely \tabularnewline
\midrule
\multicolumn{2}{p{\textwidth}}{Encourage and facilitate incremental development and fast customer feedback.  Facilitate continuous improvement by quantitative metrics, customer feedback, and the Inspect and Adapt retrospective cycle.  Engage stakeholders and provide feedback--time-boxed to an hour or less.  } \tabularnewline
\midrule
\multicolumn{2}{c}{\textbf{Quadrant 2: Scope and Requirements}} \tabularnewline
\midrule
\emph{Conflicting system requirements} & definitely \tabularnewline
\midrule
\multicolumn{2}{p{\textwidth}}{Apply enabler for exploration that provides a way for development teams to flesh out the details of requirements and design.  Continuously communicate emerging requirements and opportunities back into the program vision through product owner.  Work with stakeholders to understand the specific business targets behind the user-system interaction.  } \tabularnewline
\midrule
\emph{Continually changing project scope/objectives} & definitely \tabularnewline
\midrule
\multicolumn{2}{p{\textwidth}}{Perform system demo as near as possible to the end of the iteration.  Integrate every other iteration.  Actively participate in ongoing agreements to maintain business and development alignment as priorities and scope are inevitably changed.  } \tabularnewline
\midrule
\emph{Continually changing system requirements} & definitely \tabularnewline
\midrule
\multicolumn{2}{p{\textwidth}}{Perform system demo as near as possible to the end of the iteration.  Integrate every other iteration.  Continuously communicate emerging requirements and opportunities back into the program vision through product owner.  } \tabularnewline
\midrule
\emph{Difficulty in defining the inputs and outputs of the system} & definitely \tabularnewline
\midrule
\multicolumn{2}{p{\textwidth}}{Define PI planning's primary outputs.  Demonstrate each new feature in an end-to-end use case.  Develop a feature team that is organized around user-centered functionality. Each team is capable of delivering end-to-end value. Feature teams operate primarily with user stories, refactors, and spikes.  } \tabularnewline
\midrule
\emph{Ill-defined project goals} & definitely \tabularnewline
\midrule
\multicolumn{2}{p{\textwidth}}{Implement value stream coordination to ensure that the enterprise moves forward with each value stream in lockstep with the enterprise objectives.  Align development to business via business context, vision, and Team and Program PI Objectives.  Create a set of 'SMART' team PI objectives for each individual team with business value assigned.  } \tabularnewline
\midrule
\emph{Incorrect system requirements} & definitely \tabularnewline
\midrule
\multicolumn{2}{p{\textwidth}}{Epics and lightweight business cases.  Ensure that Epics and Enablers are reasoned and analyzed prior to reaching a Program Increment boundary, are prioritized appropriately, and have established acceptance criteria to guide a high-fidelity implementation.   Define Roadmap.  } \tabularnewline
\midrule
\emph{System requirements not adequately identified} & definitely \tabularnewline
\midrule
\multicolumn{2}{p{\textwidth}}{Epics and lightweight business cases.  Ensure that Epics and Enablers are reasoned and analyzed prior to reaching a Program Increment boundary, are prioritized appropriately, and have established acceptance criteria to guide a high-fidelity implementation.   Define Roadmap.  } \tabularnewline
\midrule
\emph{Unclear system requirements} & definitely \tabularnewline
\midrule
\multicolumn{2}{p{\textwidth}}{Epics and lightweight business cases.  Ensure that Epics and Enablers are reasoned and analyzed prior to reaching a Program Increment boundary, are prioritized appropriately, and have established acceptance criteria to guide a high-fidelity implementation.   Define Roadmap.  } \tabularnewline
\midrule
\emph{Undefined project success criteria} & definitely \tabularnewline
\midrule
\multicolumn{2}{p{\textwidth}}{Define Success criteria to validate the implementation.  Impacts the identification, success criteria, and prioritization of epics in the funnel and backlog states.  Success criteria provide a mechanism to understand progress towards the intent.  } \tabularnewline
\midrule
\emph{Users lack understanding of system capabilities and limitations} & definitely \tabularnewline
\midrule
\multicolumn{2}{p{\textwidth}}{ Perform system demos.  Integrate to illustrate a particular feature, capability, or nonfunctional requirement.  Demonstrate each new feature in an end-to-end use case.  } \tabularnewline
\midrule
\multicolumn{2}{c}{\textbf{Quadrant 3: Execution}} \tabularnewline
\midrule
\emph{Development team unfamiliar with selected development tools} & somewhat \tabularnewline
\midrule
\multicolumn{2}{p{\textwidth}}{Understand requirements for working on "technical infrastructure, tooling, and other systemic impediments".  It may be more efficient to perform an upgrade or migration at a time when there isn't a critical demo just a few days away.  Perform Daily Stand-up to understand team's status, "escalate problems, and get help from other team members".  } \tabularnewline
\midrule
\emph{Frequent conflicts among development team members} & somewhat \tabularnewline
\midrule
\multicolumn{2}{p{\textwidth}}{Vote of confidence/commitment from the entire program to these objectives.  Develop PI commitment.  Establish an agreement to determine how the work is performed for each activity type.  } \tabularnewline
\midrule
\emph{Frequent turnover within the project team} & not at all \tabularnewline
\midrule
\emph{High level of technical complexity} & definitely \tabularnewline
\midrule
\multicolumn{2}{p{\textwidth}}{Break down the architectural enabler into small enabler stories that can fit in iterations.  Perform initial exploration of epics and rank them roughly by using Weighted Shortest Job First (WSJF) to determine which epics should move to the next step for deeper exploration.  Apply Weighted Shortest Job First prioritization method for job sequencing.  } \tabularnewline
\midrule
\emph{Highly complex task being automated} & definitely \tabularnewline
\midrule
\multicolumn{2}{p{\textwidth}}{Perform initial exploration of epics and rank them roughly by using Weighted Shortest Job First (WSJF) to determine which epics should move to the next step for deeper exploration.  Apply Weighted Shortest Job First prioritization method for job sequencing.   Define Weighted Shortest Job First (WSJF).  } \tabularnewline
\midrule
\emph{Immature technology} & definitely \tabularnewline
\midrule
\multicolumn{2}{p{\textwidth}}{Work with Agile Teams that perform research spikes, create proof of concepts, mock-ups, etc.  Support technology/engineering aspects of program and value stream kanbans.  Local stories representing new functionality, refactors, defects, research spikes, and other technical debt are identified, written as enabler stories, estimated, and sequenced.  } \tabularnewline
\midrule
\emph{Inadequate estimation of project budget} & definitely \tabularnewline
\midrule
\multicolumn{2}{p{\textwidth}}{Develop Incremental implementation by keeping the epics in the portfolio backlog until there is implementation capacity available.  Avoid overhead and enables the train to make fast and local decisions within the constraints of the allocated budget.  Provide WIP limits to ensure that the teams responsible for analysis undertake into responsibly and do not create expectation for implementation or time frames that far exceed capacity and reality.  } \tabularnewline
\midrule
\emph{Inadequate estimation of project schedule} & definitely \tabularnewline
\midrule
\multicolumn{2}{p{\textwidth}}{Agile Estimating and Planning.  Adopt Agile estimating and planning by using the currency of story points.  Epic Progress Measure.  } \tabularnewline
\midrule
\emph{Inadequate estimation of required resources} & definitely \tabularnewline
\midrule
\multicolumn{2}{p{\textwidth}}{Develop Incremental implementation by keeping the epics in the portfolio backlog until there is implementation capacity available.  Provide WIP limits to ensure that the teams responsible for analysis undertake into responsibly and do not create expectation for implementation or time frames that far exceed capacity and reality.  Use capacity allocation to estimate portfolio epic based on the given knowledge of program velocities.  } \tabularnewline
\midrule
\emph{Inadequately trained development team members} & somewhat \tabularnewline
\midrule
\multicolumn{2}{p{\textwidth}}{Conduct specialized training to keep up with advancements in their respective fields.  } \tabularnewline
\midrule
\emph{Ineffective communication} & definitely \tabularnewline
\midrule
\multicolumn{2}{p{\textwidth}}{Facilitate continuous improvement by quantitative metrics, customer feedback, and the Inspect and Adapt retrospective cycle.  Establish high-bandwidth communication across all team members and stakeholders.  Perform system demo as near as possible to the end of the iteration.  } \tabularnewline
\midrule
\emph{Ineffective project manager} & not at all \tabularnewline
\midrule
\emph{Inexperienced project manager} & not at all \tabularnewline
\midrule
\emph{Inexperienced team members} & somewhat \tabularnewline
\midrule
\multicolumn{2}{p{\textwidth}}{Involve a subject matter expert in basic exploration and sizing.  Interact with analyst and subject matter experts during specification workshops.  Work with stakeholders and subject matter experts to define the epic and its potential benefits.  } \tabularnewline
\midrule
\emph{Lack of an effective project management methodology} & definitely \tabularnewline
\midrule
\multicolumn{2}{p{\textwidth}}{Coach leaders, teams, and Scrum masters in lean-Agile practices and mindsets.  Use "Agile Project Management Tools to capture stories and status, defects, test cases, estimates, actuals, assignments, burn-down chart".  } \tabularnewline
\midrule
\emph{Lack of commitment to the project among development team members} & definitely \tabularnewline
\midrule
\multicolumn{2}{p{\textwidth}}{Vote of confidence/commitment from the entire program to these objectives.  Develop PI commitment.  Limit the commitments to longer-term work, because some other item may come along that's more important than a prior commitment.  } \tabularnewline
\midrule
\emph{Lack of people skills in project leadership} & somewhat \tabularnewline
\midrule
\multicolumn{2}{p{\textwidth}}{Define Scrum Master role.  Exhibits Lean-Agile leadership.  Protects and communicates.  } \tabularnewline
\midrule
\emph{Large number of links to other systems required} & definitely \tabularnewline
\midrule
\multicolumn{2}{p{\textwidth}}{Perform initial exploration of epics and rank them roughly by using Weighted Shortest Job First (WSJF) to determine which epics should move to the next step for deeper exploration.  Apply Weighted Shortest Job First prioritization method for job sequencing.   Define Weighted Shortest Job First (WSJF).  } \tabularnewline
\midrule
\emph{Negative attitudes by development team} & somewhat \tabularnewline
\midrule
\multicolumn{2}{p{\textwidth}}{Perform Iteration Retrospective as the check step for the overall iteration.  Use "iteration retrospective to drive program level changes to process either immediately or in the Inspect and Adapt workshop".  Perform Iteration Retrospective to identify way to improve.  } \tabularnewline
\midrule
\emph{new: Ineffective collaboration} & definitely \tabularnewline
\midrule
\multicolumn{2}{p{\textwidth}}{Manage dependencies by applying extensive degree of cooperation; a common value stream backlog; implementation of new, crosscutting capabilities; additional system integration; additional roles and responsibilities; special considerations for pre-, post-, and PI planning activities; different degree and types of DevOps support.  Product manager create features in collaboration with product owner and other key stakeholders. Features are also created as a result of decomposition of epics.  Encourage the collaboration between teams and System and Solution Architects, Engineering, and User Experience designers.  } \tabularnewline
\midrule
\emph{new: Ineffective coordination} & definitely \tabularnewline
\midrule
\multicolumn{2}{p{\textwidth}}{Manage dependencies by applying extensive degree of cooperation; a common value stream backlog; implementation of new, crosscutting capabilities; additional system integration; additional roles and responsibilities; special considerations for pre-, post-, and PI planning activities; different degree and types of DevOps support.  Implement value stream coordination to ensure that the enterprise moves forward with each value stream in lockstep with the enterprise objectives.  Align development to business via business context, vision, and Team and Program PI Objectives.  } \tabularnewline
\midrule
\emph{new: Lack of trust} & definitely \tabularnewline
\midrule
\multicolumn{2}{p{\textwidth}}{Manage dependencies by applying extensive degree of cooperation; a common value stream backlog; implementation of new, crosscutting capabilities; additional system integration; additional roles and responsibilities; special considerations for pre-, post-, and PI planning activities; different degree and types of DevOps support.  Adopt SAFe's set of seven transformational patterns that can be used to move the organization to Lean-Agile Program Portfolio Management ([Decentralized decision-making], [Demand management; continuous value flow], [Lightweight; epic-only business cases], [Decentralized, rolling-wave planning], [Agile estimating and planning], [Lean-Agile budgeting and self-managing Agile Release Trains], [Objective, fact-based measure and milestones]).  Self-managing Agile Release Trains.  } \tabularnewline
\midrule
\emph{One of the largest projects attempted by the organization} & definitely \tabularnewline
\midrule
\multicolumn{2}{p{\textwidth}}{Perform initial exploration of epics and rank them roughly by using Weighted Shortest Job First (WSJF) to determine which epics should move to the next step for deeper exploration.  Apply Weighted Shortest Job First prioritization method for job sequencing.   Define Weighted Shortest Job First (WSJF).  } \tabularnewline
\midrule
\emph{Poor project planning} & definitely \tabularnewline
\midrule
\multicolumn{2}{p{\textwidth}}{Implement value stream coordination to ensure that the enterprise moves forward with each value stream in lockstep with the enterprise objectives.  Align development to business via business context, vision, and Team and Program PI Objectives.  Create a set of 'SMART' team PI objectives for each individual team with business value assigned.  } \tabularnewline
\midrule
\emph{Project affects a large number of user departments or units} & somewhat \tabularnewline
\midrule
\multicolumn{2}{p{\textwidth}}{Demonstrate each new feature in an end-to-end use case.  Develop a feature team that is organized around user-centered functionality. Each team is capable of delivering end-to-end value. Feature teams operate primarily with user stories, refactors, and spikes.  Use capabilities as end-to-end solution services that support the achievement of user goals.  } \tabularnewline
\midrule
\emph{Project involves the use of new technology} & definitely \tabularnewline
\midrule
\multicolumn{2}{p{\textwidth}}{Work with Agile Teams that perform research spikes, create proof of concepts, mock-ups, etc.  PI Features are broken into stories and placed on team backlog.  Local stories representing new functionality, refactors, defects, research spikes, and other technical debt are identified, written as enabler stories, estimated, and sequenced.  } \tabularnewline
\midrule
\emph{Project involves use of technology that has not been used in prior projects} & definitely \tabularnewline
\midrule
\multicolumn{2}{p{\textwidth}}{Work with Agile Teams that perform research spikes, create proof of concepts, mock-ups, etc.  Ensure that the demo environments are adequate to the challenge of reliably demonstrating new solution functionality.  Local stories representing new functionality, refactors, defects, research spikes, and other technical debt are identified, written as enabler stories, estimated, and sequenced.  } \tabularnewline
\midrule
\emph{Project milestones not clearly defined} & definitely \tabularnewline
\midrule
\multicolumn{2}{p{\textwidth}}{Implement value stream coordination to ensure that the enterprise moves forward with each value stream in lockstep with the enterprise objectives.  Align development to business via business context, vision, and Team and Program PI Objectives.  Create a set of 'SMART' team PI objectives for each individual team with business value assigned.  } \tabularnewline
\midrule
\emph{Project progress not monitored closely enough} & definitely \tabularnewline
\midrule
\multicolumn{2}{p{\textwidth}}{ Perform system demos.  Perform system demo as near as possible to the end of the iteration.  Demonstrate each new feature in an end-to-end use case.  } \tabularnewline
\midrule
\emph{Team members lack specialized skills required by the project} & somewhat \tabularnewline
\midrule
\multicolumn{2}{p{\textwidth}}{Involve a subject matter expert in basic exploration and sizing.  Interact with analyst and subject matter experts during specification workshops.  Work with stakeholders and subject matter experts to define the epic and its potential benefits.  } \tabularnewline
\midrule
\emph{Team members not familiar with the task(s) being automated} & definitely \tabularnewline
\midrule
\multicolumn{2}{p{\textwidth}}{Develop a feature team that is organized around user-centered functionality. Each team is capable of delivering end-to-end value. Feature teams operate primarily with user stories, refactors, and spikes.  Refine the backlog.  Involve with Agile team for short period of time.  } \tabularnewline
\midrule
\multicolumn{2}{c}{\textbf{Quadrant 4: Environment}} \tabularnewline
\midrule
\emph{Change in organizational management during the project} & not at all \tabularnewline
\midrule
\emph{Corporate politics with negative effect on project} & somewhat \tabularnewline
\midrule
\multicolumn{2}{p{\textwidth}}{Provide decision-making filters in the portfolio kanban, thereby influence the portfolio backlog.   Define Strategic Themes.  Formulate Strategic themes.  } \tabularnewline
\midrule
\emph{Dependency on outside suppliers} & definitely \tabularnewline
\midrule
\multicolumn{2}{p{\textwidth}}{Business Owner prepares to communicate the business context, including milestones and significant external dependencies, such as those of suppliers.  Work with Customers, stakeholders, and Suppliers to establish high-level Solution Intent; help establish the solution intent information models and documentation requirements.  Work with Suppliers, making sure the requirements for supplier-delivered capabilities are understood, and assist with the conceptual integration of these concerns.  } \tabularnewline
\midrule
\emph{Many external suppliers involved in the development project} & definitely \tabularnewline
\midrule
\multicolumn{2}{p{\textwidth}}{.  Business Owner prepares to communicate the business context, including milestones and significant external dependencies, such as those of suppliers.  Work with Customers, stakeholders, and Suppliers to establish high-level Solution Intent; help establish the solution intent information models and documentation requirements.  } \tabularnewline
\midrule
\emph{new: Country-specific regulations} & definitely \tabularnewline
\midrule
\multicolumn{2}{p{\textwidth}}{File patent, certify the system, audit certain regulatory requirements for economic success of product development.  Defines how the system builder's system intent must be organized, packaged, and integrated for use by the Customer to meet any compliance, certification, and other objectives.  Supports compliance and contractual obligations.  } \tabularnewline
\midrule
\emph{new: Delays caused by global distance} & somewhat \tabularnewline
\midrule
\multicolumn{2}{p{\textwidth}}{Establish high-bandwidth communication across all team members and stakeholders.  Calculate the Cost of Delay.  Manage and optimize the flow of value through the program using various tools, such as the Program and Value stream Kanbans and information radiators.  } \tabularnewline
\midrule
\emph{new: Lack of architecture-organization alignment} & somewhat \tabularnewline
\midrule
\multicolumn{2}{p{\textwidth}}{Identify dependencies and forecast cross-team and cross-ART coordination.  Create enabler features by architects or engineers to pave the architectural runway which is maintained in the program backlog alongside business features.  Create architectural enablers to pave the runway.  } \tabularnewline
\midrule
\emph{new: Lack of face-to-face interaction inhibits knowledge sharing} & definitely \tabularnewline
\midrule
\multicolumn{2}{p{\textwidth}}{Provide guidance for cross-team design and implementation synchronization.  Manage and optimize the flow of value through the program using various tools, such as the Program and Value stream Kanbans and information radiators.  RTE facilitate periodic synchronization meetings, including the ART sync at the Program level and the Value Stream (VS) sync at the value stream level.  } \tabularnewline
\midrule
\emph{new: Lack of process alignment} & definitely \tabularnewline
\midrule
\multicolumn{2}{p{\textwidth}}{Adopt decentralized, program, and team-based rolling-wave planning via routine and cadence-based PI Planning activity.  Program Portfolio Management (PPM) team continuously assesses and improves their processes using a structured, periodic Self-Assessment.  Perform a structured, root cause analysis-based, problem-solving workshop for large, systematic program-level problems.  } \tabularnewline
\midrule
\emph{new: Lack of tool/infrastructure alignment} & definitely \tabularnewline
\midrule
\multicolumn{2}{p{\textwidth}}{Manage dependencies by applying extensive degree of cooperation; a common value stream backlog; implementation of new, crosscutting capabilities; additional system integration; additional roles and responsibilities; special considerations for pre-, post-, and PI planning activities; different degree and types of DevOps support.  Create infrastructure enablers to be ready to develop, test, and integrate the initiatives.  Develop infrastructure in different levels of SAFe to support frequent or continuous integration and testing.  } \tabularnewline
\midrule
\emph{new: Unstable country/regional political/economic environment} & somewhat \tabularnewline
\midrule
\multicolumn{2}{p{\textwidth}}{Build an economically viable solution.  File patent, certify the system, audit certain regulatory requirements for economic success of product development.  Require evolution of the Customer's deployment environment for new solutions.  } \tabularnewline
\midrule
\emph{Organization undergoing restructuring during the project} & not at all \tabularnewline
\midrule
\emph{Resources shifted from the project due to changes in organizational priorities} & somewhat \tabularnewline
\midrule
\multicolumn{2}{p{\textwidth}}{Apply WSJF to prioritize the items in the Value stream and Program backlogs.  PI cadence offers time box experimenting approach which later help to produce the desired and long term results.  Apply WSJF prioritization to compare business and enabler epics against each other.  } \tabularnewline
\midrule
\emph{Unstable organizational environment} & not at all \tabularnewline
 
  \bottomrule
\end{longtable}
\end{singlespace}
\end{tiny}

\end{document}